\documentclass[12pt]{article}

\usepackage{newtxtext,newtxmath}

\usepackage{graphicx}

\usepackage[letterpaper,margin=1in]{geometry}

\renewenvironment{abstract}
	{\quotation}
	{\endquotation}

\date{}

\makeatletter
\renewcommand{\fnum@figure}{\textbf{Figure \thefigure}}
\renewcommand{\fnum@table}{\textbf{Table \thetable}}
\makeatother

\usepackage{scicite}

\usepackage{url}

\def\scititle{
	Generative AI models capture realistic sea-ice evolution from days to decades
}
\title{\bfseries \boldmath \scititle}

\author{
	Tobias~Sebastian~Finn,$^{1\ast}$
	Marc~Bocquet,$^{1}$
	Pierre~Rampal$^{2}$\and
    Charlotte~Durand$^{2}$\and
    Flavia~Porro$^{3}$\and
    Alban~Farchi$^{4}$\and
    Alberto~Carrassi$^{3}$\and
	\small$^{1}$CEREA, ENPC, EDF R\&D, Institut Polytechnique de Paris, Marne-la-Vallée, 77455, France.\and
	\small$^{2}$Institut des Geosciences de l'Environnement, CNRS, Grenoble, France.\and
	\small$^{3}$Dept. of Physics and Astronomy “Augusto Righi”, University of Bologna, Bologna, Italy.\and
	\small$^{4}$ECMWF, European Centre for Medium-Range Weather Forecast, Reading, RG2 9AX, United Kingdom.\and
	\small$^\ast$Corresponding author. Email: tobias.finn@enpc.fr
}

\begin{document} 

\maketitle

\begin{abstract} \bfseries \boldmath
    Sea ice plays an important role in stabilising the Earth system.
    Yet, representing its dynamics remains a major challenge for models, as the underlying processes are scale-invariant and highly anisotropic.
    This poses a dilemma: physics-based models that faithfully reproduce the observed dynamics are computationally costly, while efficient AI models sacrifice realism.
    Here, to resolve this dilemma, we introduce \textit{GenSIM}, the first generative AI model to predict the evolution of the full Arctic sea-ice state at 12-hour increments.
    Trained for sub-daily forecasting on 20 years of sea-ice--ocean simulation data, GenSIM makes realistic predictions for 30 years, while reproducing the dynamical properties of sea ice with its leads and ridges and capturing long-term trends in the sea-ice volume.
    Notably, although solely driven by atmospheric reanalysis, GenSIM implicitly learns hidden signatures of multi-year ice–ocean interaction.
    Therefore, generative AI can extrapolate from sub-daily forecasts to decadal simulations, while retaining physical consistency.
\end{abstract}

\noindent

Sea ice stabilises the Earth system by regulating thermohaline circulation and modulating albedo.
Its mechanical dynamics arise from the transition of a fluid flow to a fractured, discrete continuum, marked by scale invariance, multi fractality, and anisotropy \cite{marsan_scale_2004, rampal_scaling_2008}.
These processes are captured in physics-based numerical models through rheologies that describe how the sea ice deforms under external forcings such as winds, currents, and waves \cite{hibler_dynamic_1979}.
Recent sea-ice models with brittle rheologies \cite{dansereau_maxwell_2016,rampal_nextsim_2016,brodeau_implementation_2024} use a \textit{damage} variable to parametrise subgrid fracturing and to reproduce the multi-scale dynamics observed in high-resolution satellite data \cite{rampal_multifractal_2019, olason_new_2022, bouchat_sea_2022}.
Although this increases their realism, these models typically require advanced numerical techniques such as Lagrangian finite elements \cite{rampal_nextsim_2016}, which raise their computational costs and complicate ensemble forecasts and long-term simulations \cite{samake_parallel_2017, brodeau_implementation_2024}.

This highlights a long-standing dilemma: sea-ice models that faithfully reproduce the observed dynamics require either kilometre-scale resolutions or brittle rheologies \cite{bouchat_sea_2022}, whereas computationally efficient models sacrifice realism.
In recent years, computational gains are offered by surrogate models that leverage deep neural network.
For sea ice, most approaches so far have focused on univariate predictions of sea-ice thickness \cite{durand_datadriven_2024} or sea-ice concentration and extent \cite{andersson_seasonal_2021, liu_daily_2021, kvanum_developing_2024}, missing the full picture of the sea-ice state.
Crucially, trained as deterministic models, they suffer from a loss of small-scale information and smoothing \cite{durand_datadriven_2024, kochkov_neural_2024}, and reduced physical consistency \cite{bonavita_limitations_2024} compared to physics-based models.
Moreover, their auto-regressive predictions risk an error accumulation \cite{parthipan_defining_2024} leading to instabilities in long-term simulations \cite{chattopadhyay_challenges_2024}.

Generative diffusion and flow matching models \cite{sohl-dickstein_deep_2015, song_scorebased_2021,liu_flow_2022,lipman_flow_2023,albergo_stochastic_2023} remedy these limitations by learning to produce diverse state realisations.
Conditioned on the current state of the system and external forcings, this enables applications from ensemble generation \cite{li_seeds_2023, finn_representation_2024, meuer_latent_2024}, over weather prediction \cite{price_probabilistic_2025} and downscaling \cite{mardani_residual_2024}, to climate emulation \cite{cachay_probabilistic_2024,brenowitz_climate_2025}.
When applied to sea-ice predictions, these models can refine sea-ice concentration forecasts \cite{xu_icediff_2024}, generate sea-ice states \cite{finn_diffusion_2024}, and predict sea ice on regional scales \cite{finn_generative_2024}.
Nonetheless, their computational demands hamper adoption to global-scale predictions, and their iterative generation remains unconstrained by physical bounds such as non-negativity of the sea-ice thickness.

We resolve this dilemma with GenSIM, the first generative AI-based sea-ice model that predicts the full sea-ice state across spatial scales from $10\,\text{km}$ to the Arctic basin and temporal scales from $12\,\text{hours}$ to multi-decadal climate trends.
Instantiated as an auto-regressive flow matching model, GenSIM utilises a scale-aware transformer architecture and exploits physical localisation by domain decomposition for efficient predictions, being up to 50 times faster than the reference physics-based neXtSIM-OPA configuration \cite{boutin_arctic_2023}.
Trained on 20 years of neXtSIM-OPA output, GenSIM’s novel censored loss function enforces physical bounds during training.
These advances allow GenSIM to make realistic predictions over 30 years, reproducing multifractal brittle-like short-term dynamics, the decades-long sea-ice decline, and the recent hiatus due to internal variability \cite{england_minimal_2025}, as validated against physics-based models and satellite observations
Although forced solely by the atmosphere, GenSIM learns to extrapolate by implicitly encoding the multi-year interactions between sea ice and ocean.
Trained for sub-daily forecasting, GenSIM can robustly perform multi-decadal predictions and capture multi-scale physical processes, thereby effectively removing the common inability of simultaneously learn multiple scales in AI-based Earth system models.

\subsection*{Sea-ice modelling with GenSIM}\label{sec:gensim}

GenSIM is a sea-ice model developed to generate 12-hour forecasts of sea-ice thickness, concentration, damage, drift components, and snow-on-ice thickness over the full Arctic basin.
The model is driven by its initial conditions and external forcings from the atmosphere: the 2-metre temperature and specific humidity, and 10-metre zonal and meridional wind.
To extend the forecast window beyond the trained 12~hours, GenSIM is implemented as an autoregressive system, where each prediction serves as initial condition for the subsequent 12~hours, similar to prediction models for the atmosphere \cite{price_probabilistic_2025}.
Natively operating on a curvilinear $0.25^{\circ}$ mesh (approximately~12 km resolution), GenSIM makes 12~hour forecasts in a few seconds and needs below $4\,\text{GB}$ memory, which enables large ensembles and decadal-long simulations on a single consumer-grade GPU (here, an Nvidia RTX A6000 Ampere with $48\,\text{GB}$ VRAM).

\begin{figure}[ht]
    \centering
    \includegraphics[width=0.6\textwidth]{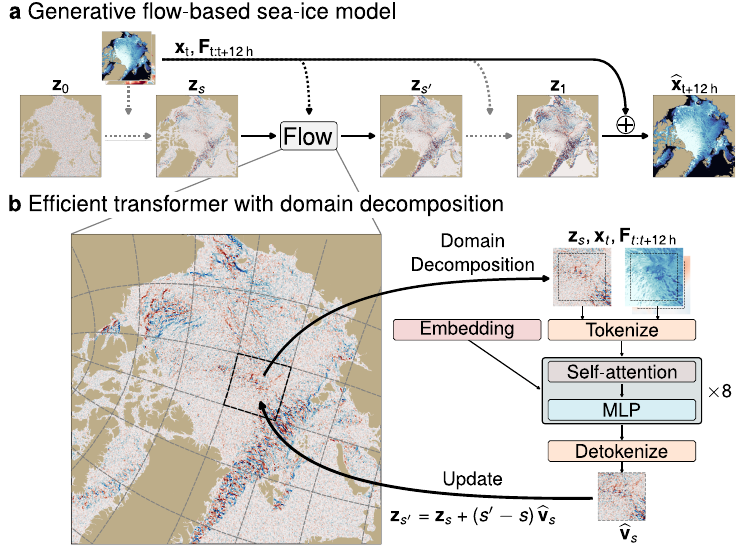}
    \caption{
        \textbf{Schematic of the efficient generative AI-based sea-ice model {GenSIM}.}
        Based on flow matching, GenSIM jointly predicts all important sea-ice states for a lead time of 12~hours.
        (a) Conditioned on initial states $\mathbf{x}_{t}$ and external atmospheric forcings $\mathbf{F}_{t:t+12\,\text{h}}$, the model iteratively generates a prediction $\widehat{\mathbf{x}}_{t+12\,\text{h}}$ 12~hours later.
        (b) Within an iterative flow step, at arbitrary pseudo time $s \in [0, 1]$, the full Arctic domain is decomposed into overlapping subdomains and processed in parallel with a scale-aware transformer, which predicts a velocity, used to update the fields to the next pseudo tome step $s' \ge s$.
    }\label{fig:fig_schematic}
\end{figure}

Instantiated as flow matching model \cite{lipman_flow_2023} and conditioned on the initial conditions, $\mathbf{x}_{t}$, and forcings, $\mathbf{F}_{t:t+12\,\text{h}}$, GenSIM produces forecasts by iteratively mapping samples from a Gaussian distribution, $\mathbf{z}_{0}$, to tendency fields, $\mathbf{z}_{1}$, as schematically depicted in Fig.~\ref{fig:fig_schematic}a.
Added to the initial conditions, these tendencies yield the 12-hour forecast, $\widehat{\mathbf{x}}_{t + 12\,\text{h}}$.
Crucially, varying the initial noise sample allows GenSIM to produce an ensemble of forecasts from identical initial conditions and forcings, representing model uncertainty.

Sea ice is physically bounded (e.g., non-negativity of the thickness).
To follow these bounds, we introduce censored flow matching based on variational flow matching \cite{eijkelboom_variational_2024} and censored Gaussian distributions \cite{finn_diffusion_2024}, effectively constraining the generative flow.
The resulting loss function adjusts the usual mean-squared error when encountering bounded data.
There, the loss drives the model towards more extreme values to increase the probability that the bound is exceeded as exposed in the mathematical framework (Supplementary Text).

Integrated in pseudo time from $s=0$ (pure noise) to $s=1$ (clean tendency), the flow and the underlying dynamical system are defined by its generative velocity, $\widehat{\mathbf{v}}_{s}$.
A trained neural network approximates the generative velocity to propagate the state to the next pseudo-time step, $\mathbf{z}_{s'}$, given the intermediate states $\mathbf{z}_{s}$, the initial conditions, forcings, and any relevant embeddings as derived from the pseudo time, augmentation labels, and the used resolution.

Considering the similarities between this generative flow and traditional numerical methods \cite{finn_representation_2024}, and leveraging the localised physical sensitivity of the tendencies, we use a domain decomposition strategy \cite{zhang_diffcollage_2023, brenowitz_climate_2025} to enhance computational efficiency.
Instead of processing global fields, the neural network predicts velocities within decomposed subdomains (Fig.~\ref{fig:fig_schematic}b).
To foster communication between subdomains when integrating the flow, the inputted subdomains overlap, while the predicted velocity itself remains local.
This decomposition enables for fully parallel processing and a linear scaling with the number of global grid points, i.e., the number of subdomains is doubled when the number of global grid points is doubled.

The neural network adopts a scale-aware, mesh-independent transformer architecture, designed to explicitly encode spatial interactions and to accommodate masked land cells.
Initially, through a linear tokenizer with a $2\times2$ patching \cite{peebles_scalable_2023}, the network converts the input into tokens.
These tokens are processed by a series of transformer blocks: self-attention blocks capture spatial dependencies with a learnable localisation that depends on relative distances \cite{vaswani_attention_2017,heo_rotary_2024}, while multi-layer perceptrons (MLPs) non-linearly combine the resulting spatial representations.
The embedding modulates thereby the inputs to the attention and MLP blocks.
After $8$ transformer blocks, the tokens are mapped to local velocities via a detokenisation layer.

We trained the model on 20 years (1995--2014) from a long simulation with the coupled sea-ice--ocean neXtSIM-OPA system \cite{boutin_arctic_2023} and forced by the atmosphere with the ERA5 reanalysis \cite{hersbach_era5_2020}.
To leverage the scale invariance of sea ice and to enhance mesh and resolution independence, we augmented the data by averaging over randomly drawn kernel sizes up to $8\times8$.
The domain decomposition is implemented during training by randomly slicing subdomains from the pan-Arctic domain to match the sampling subdomain size ($80\times80$ grid points, including $8$ points overlap in each direction).
This efficiently enabled us to train the model on eight GPUs only and cut the training cost to around two days \cite{wang_patch_2024}.
Finally, we employed distribution augmentation \cite{jun_distribution_2020} with random flipping and rotations to promote the model's ability to generalise.

Before applying GenSIM to decadal long-term simulations, we evaluated its performance in emulating neXtSIM-OPA on held-out data from 2016--2018 (Supplementary Text, figure~\ref{fig:main_short_forecast},~figure~\ref{fig:app_scorecard_a}, and~figure~\ref{fig:app_scorecard_b}).
GenSIM outperforms a deterministic baseline for this period, reducing in its mean the root-mean-squared error by around $20\,\%$, matching the energy spectrum from neXtSIM-OPA for individual members, and improving the representation of the marginal ice zone.
Additionally, GenSIM's ensemble exhibits typical spread-skill ratios around $0.8$, remaining just slightly underdispersive.
These results make GenSIM a major step towards realistically emulating physics-based sea-ice models and improving upon deterministic models.

\subsection*{Realistic decade-long simulations}

To analyse the long-term behaviour of GenSIM, we perform a 30 years long simulation with a single ensemble member, similar to the neXtSIM-OPA simulation that was used to train our model.
Starting from the same initial conditions as neXtSIM-OPA and forced by ERA5, GenSIM is run autoregressively over the period 1995--2024 with the first five years used as spin-up and the last ten years as extrapolation period.

\begin{figure}[ht!]
    \centering
    \includegraphics[width=0.6\textwidth]{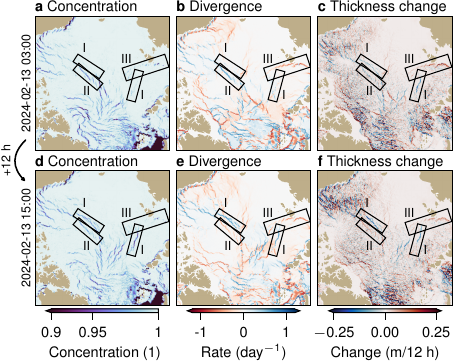}
    \caption{
        \textbf{Physical consistency of GenSIM.}
        Snapshots from a 30-year-long GenSIM simulation for (a--c) 2024-02-13 03:00 UTC and (d--f) 2024-02-13 15:00 UTC with shown sea-ice concentration (a\&d), divergence rate (b\&e), and change in the sea-ice thickness within 12 hours (c\&f).
        Regions of interest for the (I) opening of leads, (II) healing, and (III) ridging are shown in black boxes.
    }\label{fig:consistency}
\end{figure}

Driven by the mechanical response to external atmospheric forcings, sea ice exhibits a cycle of processes from opening of leads, over healing, to ridging.
Even though GenSIM has no processes explicitly encoded, it successfully captures their impact in space and time (Fig.~\ref{fig:consistency}): (I) a small reduction in the sea-ice concentration suggests a weakening of sea ice, making it more vulnerable against stress induced by atmospheric winds.
The external stress results into a divergent deformation, which leads to ice thinning.
This thinning amplifies the response such that larger leads are opened.
(II) After the external forcing is removed or when the stress is sufficiently decreased, opened leads are healed, resulting in a subsequent increase in concentration.
(III) When linked to dynamical ridging, this healing is often accompanied by convergence, where sea ice is piled up, resulting into a strong thickening of ice.
The fact that we can track the impact of these processes in the predictions suggests that GenSIM remains physically consistent even after nearly 30 years of simulation.

This physical consistency demonstrates the strengths of generative flow models in learning high-dimensional correlations and mechanical processes in space and across variables.
Although the tendency fields are independently generated from Gaussian noise for each prediction time, they are linked in time through auto-regressive conditioning.
Moreover, as the tendencies are added up over time, GenSIM yields temporal consistency as required to exhibit such realistic cycles of damaging and healing.
Consequently, the auto-regressive application of generative flows by predicting tendencies is likely key to achieve this physical consistency in space and time.

Trained to predict the sea ice over 12~hours with external atmospheric forcings, GenSIM constrains the short-term dynamics well, producing realistic scale-invariant and anisotropic brittle-like features (Supplementary Text, ~figure~\ref{fig:deformation}).
Long-term dynamics and especially the thermodynamics are however difficult to learn due to their diminishing signal-to-noise ratio on sub-daily scales and, importantly, the missing ocean forcings.

To benchmark the thermodynamical properties, we compare GenSIM's simulated sea-ice volume to the original neXtSIM-OPA simulation and the PIOMAS reanalysis product \cite{zhang_modeling_2003}.
As observational references, we additionally use satellite products derived from CryoSat2, and Soil Moisture and Ocean Salinity (SMOS), called CS2/SMOS \cite{ricker_weekly_2017}, as well as IceSAT2 observations \cite{petty_winter_2023}.

\begin{figure}[ht!]
    \centering
    \includegraphics[width=0.8\textwidth]{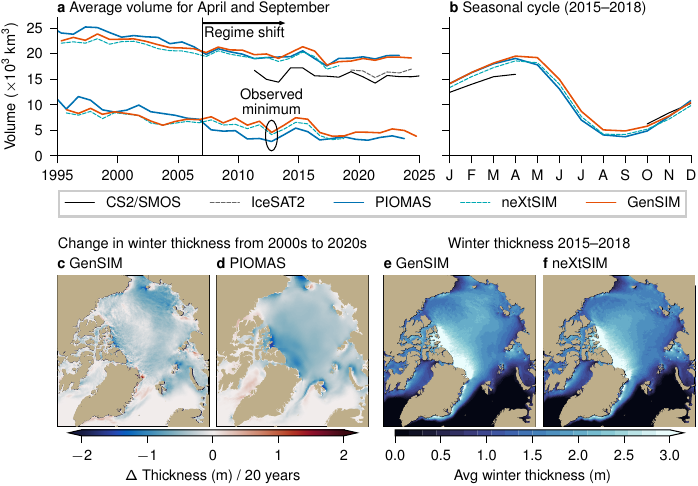}
    \caption{
        \textbf{Overview over GenSIM's long-term projection qualities.}
        (a) The simulated monthly averaged sea-ice volume in April (maximum) and September (minimum) for GenSIM, neXtSIM-OPA, PIOMAS, the merged CS2/SMOS product, and IceSAT2.
        Note that CS2/SMOS and IceSAT2 are unavailable for September.
        (b) The seasonal cycle of the average volume averaged for the period 2015--2018 with the same colouring as for (a).
        (c, d) The change in the winter thickness between 2020-2024 to 2000--2004 for GenSIM and PIOMAS.
        (e, f) The winter sea-ice thickness averaged between 2015--2018 for GenSIM and neXtSIM.
        We define the winter season as January to April.
    }\label{fig:main_long_forecast}
\end{figure}

Remarkably, GenSIM with its autoregressive steps is stable over 30~years, even in a warming climate (Fig.~ \ref{fig:main_long_forecast}a): the simulation successfully reproduces the observed seasonal cycle and closely aligns with the results from neXtSIM and PIOMAS (Fig.~\ref{fig:main_long_forecast}b).
GenSIM shows a small positive bias, especially during summer (September), attenuating the Arctic sea-ice regime shift post-2007 \cite{sumata_regime_2023}.
This bias is primarily due to an overestimation of the sea‑ice area during summer (Supplementary Text,~figure~\ref{fig:app_thickness_area}).
As a consequence, GenSIM underestimates the climate change impact for the period 2000--2018 with a linear trend of $187\pm10\,\text{km}^3 \,\text{year}^{-1}$ compared to $215\pm9\,\text{km}^3\,\text{year}^{-1}$ for neXtSIM and $329\pm12\,\text{km}^3\,\text{year}^{-1}$ for PIOMAS (only uncertainties from the regresssion itself are included).
Overestimating the observed minimum in 2012, GenSIM and neXtSIM reach their lowest September estimates in 2017 and 2018.
Nonetheless, GenSIM successfully predicts the more recent slowing of sea-ice decline (trend estimate: $104\pm29\,\text{km}^3 \,\text{year}^{-1}$ for the period 2015--2024) due to Earth system's internal variability, consistent with satellite observations and reanalysis datasets \cite{england_minimal_2025} (Supplementary Text,~figure~\ref{fig:app_extent}).

By comparing the spatial distribution in averaged winter sea-ice thickness for the periods 2000--2004 and 2020--2024 for GenSIM and PIOMAS, the underestimation of the trend becomes especially evident in the central Arctic (Fig.~\ref{fig:main_long_forecast}c and d).
Yet, GenSIM's spatial thickness distribution closely follows neXtSIM's for 2015--2018 (Fig.~\ref{fig:main_long_forecast}e and f), despite a persisting small bias towards thicker sea ice.
Even though GenSIM is exposed to biases in its volume estimates and trends, these results suggest that GenSIM has learned the seasonal character of sea ice and the long-term impact of climate change, as also seen in the diagnosed sea-ice age (Supplementary Text,~figure~\ref{fig:app_ice_age}).

The neXtSIM-OPA system \cite{boutin_arctic_2023} explicitly accounts for sea-ice--ocean interactions by coupling the sea-ice model neXtSIM \cite{rampal_nextsim_2016,olason_new_2022} to the ocean component of the NEMO modelling framework \cite{madec_nemo_2023}.
Being solely forced by the atmosphere, GenSIM contrastingly neglects an explicit representation of these interactions.
Nevertheless, GenSIM captures the long-term decline in sea-ice volume as driven by thermodynamics, with only a small bias compared to the neXtSIM-OPA simulation.
Consequently, we need to understand how GenSIM reproduces these long-term impacts so well.

\begin{figure}[ht!]
    \centering
    \includegraphics[width=1\textwidth]{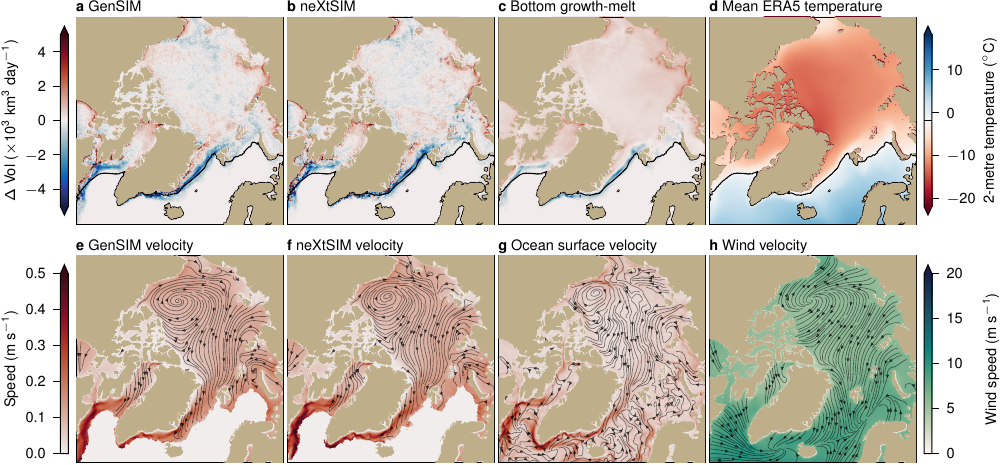}
    \caption{
        \textbf{GenSIM exhibits long-term ocean-like patterns.}
        (a--d) Volume change within 12 hours for (a) GenSIM, (b) neXtSIM, (c) net bottom growth - bottom melt from neXtSIM, and (d) average 2-metre temperature from the ERA5 reanalysis, averaged for 2015--2018.
        The black contour line indicate the $-1.8\,^{\circ}\text{C}$ isotherm from the averaged 2-metre temperature.
        The dynamical component due to advection has been removed from the changes in (a) and (b).
        (e--h) Speed (background) and streamlines (black lines) for (e) sea ice in GenSIM, (f) sea ice in neXtSIM, (g) ocean surface from the neXtSIM-OPA simulation, and (h) the 10-metre wind from the ERA5 reanalysis, averaged for 2015--2018.
    }\label{fig:long_term_dynamics}
\end{figure}

Averaged for 2015--2018, GenSIM's large-scale patterns in the thermodynamics and dynamics closely resemble those of the coupled neXtSIM-OPA simulation (Fig.~\ref{fig:long_term_dynamics}).
Thermodynamically, GenSIM exhibits the same regions of ice growth and melting as neXtSIM (Fig.~\ref{fig:long_term_dynamics}, a and b): major growth occurs in the Laptev sea and Kara sea, as well as the Canadian archipelago and central Arctic, while the strongest melt is found around Greenland, in the Hudson bay, the Barents sea, and the Chukchi sea.
The melt-dominated regions are linked to bottom melting (Fig. \ref{fig:long_term_dynamics}, c), as sea ice is exposed to warmer water from the Atlantic or Pacific ocean therein.

These spatial patterns by GenSIM cannot be explained by the average 2-metre temperature alone (Fig. \ref{fig:long_term_dynamics}, d).
Instead, GenSIM infers them from combining all atmospheric forcings together with the land mask, which allows it to reproduce signatures that in nature arise from ocean–ice interactions.
As a result, the neural network identifies the same key melting zones and yields melting rates comparable to those of neXtSIM-OPA.
GenSIM implicitly captures the long-term thermodynamical impact of ocean-ice interactions.

Although the sea-ice dynamics are dominated by the atmosphere on daily scales \cite{mohammadi-aragh_predictability_2018}, the dynamics of sea ice and ocean remain aligned on year-long scales, as they are both driven by the general circulation from the atmosphere.
Consequently, we find an agreement in the sea-ice drift and the ocean surface velocity in coupled sea-ice--ocean simulations (Fig. \ref{fig:long_term_dynamics}, f and g): sea ice circulates in front of the Alaskan and Canadian coast due to the Beaufort gyre, while the transpolar drift transports sea ice from growing zones to melting zones through the Fram strait.
GenSIM's sea-ice drift matches these major ocean currents in the Arctic basin (Fig. \ref{fig:long_term_dynamics}, e), even though the effective Beaufort gyre is displaced.
As only forcing available for sea-ice motion, the general atmospheric circulation (Fig. \ref{fig:long_term_dynamics}, h) determines GenSIM's overall smoother patterns.
Remaining differences can be attributed to the mechanical response of sea ice on shorter time scales and to other indirect factors, such as the sea-ice concentration.
Once again, GenSIM has apparently learned the long-term link between coupled sea-ice--ocean dynamics and atmospheric forcings by extracting spatial patterns from these forcings.

Without direct access to external oceanic forcings, GenSIM has nonetheless learned the long-term interplay between sea ice and ocean present in the coupled neXtSIM-OPA simulation.
The learning of this interplay emerges from patterns that are encoded in the atmospheric forcings and their relationship to the land mask.
Consequently, although trained solely trained to predict for a 12-hour lead time, GenSIM correctly infers slow multi-year effects and uncovers hidden influences from the ocean, even when their signal-to-noise ratio is diminishing for the training task.
These results explain how GenSIM can match its long-term evolution to advanced physics-based models and satellite observations in a warming climate.

\subsection*{Consequences for AI-based Earth system models}

Our introduced generative sea-ice model, GenSIM, improves upon deterministic models by realistically capturing the full image of the Arctic sea ice: all key indicators are predicted with an unprecedented physical consistency across spatial scales from 10 km to the Arctic basin and temporal scales from 12 hours to multi-decadal climate trends.
These findings pave a way towards the use of generative diffusion and flow matching models for surrogates of other Earth system components than the atmosphere.
Furthermore, large ensembles produced by such generative models can especially be useful for sea-ice data assimilation \cite{chen_multivariate_2024}, which has been hindered by high computational costs and the unavailability of adjoints \cite{durand_fourdimensional_2025}.
Hence, generative surrogate models unlock a large potential to improve sea-ice forecasting in the future.

We built upon the observation that generative flow models work similarly to physics-based numerical models: they integrate a dynamical system that is defined by the neural network.
This similarity allows us to bridge methods used for physics-based models to generative surrogates.
Here, we use domain decomposition to efficiently apply such surrogates at global scale.
Combined with the employed transformer architecture, the surrogate is flexible in its use with applications from regional to pan-Arctic scales, generalising across scales (Supplementary Text,~figure~\ref{fig:app_coarse}).
Thanks to this flexibility, we can train generative surrogates across heterogenous datasets and resolutions \cite{brenowitz_climate_2025}, potentially further boosting their performance.

GenSIM realistically simulates decade-long climate conditions, suggesting that generative surrogates can generalise sub-daily training objectives to long-term projections.
We partially attribute its stability to GenSIM's physical consistency.
The underlying physics manifests in the system's states as evolving correlations across space, time, and between variables.
Consequently, this physical consistency is a strength of generative diffusion and flow matching models, which excel in learning high-dimensional, multi-scale correlations.
Moreover, on longer time scales, sea ice is dominantly driven by the thermodynamics and ice-ocean interactions.
With both well-captured when atmospheric forcings are provided, this further constrains and stabilises the model.
Therefore, the long-term stability is likely an inherently emerging capability of generative flows employed as sea-ice models.

The successful extrapolation to long-term evolutions from short-term tasks emerges from learning the influence of slow components on the faster, targetted, time scales.
This allows us for example to unveil the interactions of the ice with the ocean, although GenSIM is solely forced by the atmosphere.
Consequently, generative flow models can retrieve hidden dynamics in the data, even for weak signal-to-noise ratios.

The most common strategy for Earth system modelling is to have models for each Earth system component that interact through coupling.
Yet, this strategy can fail for AI-based component models: as we have demonstrated, the dynamics of these models show an implicit entanglement between components, here sea ice and ocean. 
This makes it challenging to isolate the dynamics of a single component due to external forcings from other components.
Therefore, our findings suggest the need to rethink how to do coupled Earth system modelling, when using AI-based models.

\newpage


\clearpage 

%
\bibliography{literature} 

@misc{albergo_stochastic_2023,
  title = {Stochastic {{Interpolants}}: {{A Unifying Framework}} for {{Flows}} and {{Diffusions}}},
  shorttitle = {Stochastic {{Interpolants}}},
  author = {Albergo, Michael S. and Boffi, Nicholas M. and {Vanden-Eijnden}, Eric},
  year = 2023,
  month = nov,
  number = {arXiv:2303.08797},
  eprint = {2303.08797},
  primaryclass = {cond-mat},
  publisher = {arXiv},
  doi = {10.48550/arXiv.2303.08797},
  urldate = {2024-07-17},
  archiveprefix = {arXiv}
}

@article{amemiya_regression_1973,
  title = {Regression {{Analysis}} When the {{Dependent Variable Is Truncated Normal}}},
  author = {Amemiya, Takeshi},
  year = 1973,
  month = nov,
  journal = {Econometrica},
  volume = {41},
  number = {6},
  eprint = {1914031},
  eprinttype = {jstor},
  pages = {997},
  publisher = {[Wiley, Econometric Society]},
  issn = {00129682},
  doi = {10.2307/1914031},
  urldate = {2025-04-24}
}

@article{andersson_seasonal_2021,
  title = {Seasonal {{Arctic}} Sea Ice Forecasting with Probabilistic Deep Learning},
  author = {Andersson, Tom R. and Hosking, J. Scott and {P{\'e}rez-Ortiz}, Mar{\'i}a and Paige, Brooks and Elliott, Andrew and Russell, Chris and Law, Stephen and Jones, Daniel C. and Wilkinson, Jeremy and Phillips, Tony and Byrne, James and Tietsche, Steffen and Sarojini, Beena Balan and {Blanchard-Wrigglesworth}, Eduardo and Aksenov, Yevgeny and Downie, Rod and Shuckburgh, Emily},
  year = 2021,
  month = aug,
  journal = {Nature Communications},
  volume = {12},
  number = {1},
  pages = {5124},
  publisher = {Nature Publishing Group},
  issn = {2041-1723},
  doi = {10.1038/s41467-021-25257-4},
  urldate = {2022-06-27},
  copyright = {2021 The Author(s)},
  langid = {english}
}

@misc{batzolis_conditional_2021,
  title = {Conditional {{Image Generation}} with {{Score-Based Diffusion Models}}},
  author = {Batzolis, Georgios and Stanczuk, Jan and Sch{\"o}nlieb, Carola-Bibiane and Etmann, Christian},
  year = 2021,
  month = nov,
  number = {arXiv:2111.13606},
  eprint = {2111.13606},
  primaryclass = {cs, stat},
  publisher = {arXiv},
  doi = {10.48550/arXiv.2111.13606},
  urldate = {2024-02-16},
  archiveprefix = {arXiv}
}

@article{bonavita_limitations_2024,
  title = {On {{Some Limitations}} of {{Current Machine Learning Weather Prediction Models}}},
  author = {Bonavita, Massimo},
  year = 2024,
  month = jun,
  journal = {Geophysical Research Letters},
  volume = {51},
  number = {12},
  pages = {e2023GL107377},
  issn = {0094-8276, 1944-8007},
  doi = {10.1029/2023GL107377},
  urldate = {2024-09-26},
  copyright = {\copyright{} 2024 ECMWF.},
  langid = {english}
}

@article{bouchat_reassessing_2020,
  title = {Reassessing the {{Quality}} of {{Sea-Ice Deformation Estimates Derived From}} the {{RADARSAT Geophysical Processor System}} and {{Its Impact}} on the {{Spatiotemporal Scaling Statistics}}},
  author = {Bouchat, Am{\'e}lie and Tremblay, Bruno},
  year = 2020,
  journal = {Journal of Geophysical Research: Oceans},
  volume = {125},
  number = {8},
  pages = {e2019JC015944},
  issn = {2169-9291},
  doi = {10.1029/2019JC015944},
  urldate = {2025-06-13},
  copyright = {\copyright 2020. American Geophysical Union. All Rights Reserved.},
  langid = {english}
}

@article{bouchat_sea_2022,
  title = {Sea {{Ice Rheology Experiment}} ({{SIREx}}): 1. {{Scaling}} and {{Statistical Properties}} of {{Sea}}-{{Ice Deformation Fields}}},
  shorttitle = {Sea {{Ice Rheology Experiment}} ({{SIREx}})},
  author = {Bouchat, Am{\'e}lie and Hutter, Nils and Chanut, J{\'e}r{\^o}me and Dupont, Fr{\'e}d{\'e}ric and Dukhovskoy, Dmitry and Garric, Gilles and Lee, Younjoo J. and Lemieux, Jean-Fran{\c c}ois and Lique, Camille and Losch, Martin and Maslowski, Wieslaw and Myers, Paul G. and {\'O}lason, Einar and Rampal, Pierre and Rasmussen, Till and Talandier, Claude and Tremblay, Bruno and Wang, Qiang},
  year = 2022,
  month = apr,
  journal = {Journal of Geophysical Research: Oceans},
  volume = {127},
  number = {4},
  pages = {e2021JC017667},
  issn = {2169-9275, 2169-9291},
  doi = {10.1029/2021JC017667},
  urldate = {2022-09-29},
  langid = {english}
}

@article{boutin_arctic_2023,
  title = {Arctic Sea Ice Mass Balance in a New Coupled Ice--Ocean Model Using a Brittle Rheology Framework},
  author = {Boutin, Guillaume and {\'O}lason, Einar and Rampal, Pierre and Regan, Heather and Lique, Camille and Talandier, Claude and Brodeau, Laurent and Ricker, Robert},
  year = 2023,
  month = feb,
  journal = {The Cryosphere},
  volume = {17},
  number = {2},
  pages = {617--638},
  publisher = {Copernicus GmbH},
  issn = {1994-0424},
  doi = {10.5194/tc-17-617-2023},
  urldate = {2023-12-20},
  langid = {english}
}

@article{braithwaite_positive_1995,
  title = {Positive Degree-Day Factors for Ablation on the Greenland Ice Sheet Studied by Energy-Balance Modelling},
  author = {Braithwaite, Roger J.},
  year = 1995,
  month = jan,
  journal = {Journal of Glaciology},
  volume = {41},
  number = {137},
  pages = {153--160},
  issn = {0022-1430, 1727-5652},
  doi = {10.3189/S0022143000017846},
  urldate = {2025-08-13},
  langid = {english}
}

@misc{brenowitz_climate_2025,
  title = {Climate in a Bottle: {{Towards}} a Generative Foundation Model for the Kilometer-Scale Global Atmosphere},
  shorttitle = {Climate in a {{Bottle}}},
  author = {Brenowitz, Noah D. and Ge, Tao and Subramaniam, Akshay and Gupta, Aayush and Hall, David M. and Mardani, Morteza and Vahdat, Arash and Kashinath, Karthik and Pritchard, Michael S.},
  year = 2025,
  month = may,
  urldate = {2025-06-04},
  langid = {english}
}

@article{brodeau_implementation_2024,
  title = {Implementation of a Brittle Sea Ice Rheology in an {{Eulerian}}, Finite-Difference, {{C-grid}} Modeling Framework: Impact on the Simulated Deformation of Sea Ice in the {{Arctic}}},
  shorttitle = {Implementation of a Brittle Sea Ice Rheology in an {{Eulerian}}, Finite-Difference, {{C-grid}} Modeling Framework},
  author = {Brodeau, Laurent and Rampal, Pierre and {\'O}lason, Einar and Dansereau, V{\'e}ronique},
  year = 2024,
  month = aug,
  journal = {Geoscientific Model Development},
  volume = {17},
  number = {15},
  pages = {6051--6082},
  publisher = {Copernicus GmbH},
  issn = {1991-959X},
  doi = {10.5194/gmd-17-6051-2024},
  urldate = {2025-08-07},
  langid = {english}
}

@misc{cachay_probabilistic_2024,
  title = {Probabilistic Emulation of a Global Climate Model with Spherical {{DYffusion}}},
  author = {Cachay, Salva R{\"u}hling and Henn, Brian and {Watt-Meyer}, Oliver and Bretherton, Christopher S. and Yu, Rose},
  year = 2024,
  month = nov,
  number = {arXiv:2406.14798},
  eprint = {2406.14798},
  primaryclass = {cs},
  publisher = {arXiv},
  doi = {10.48550/arXiv.2406.14798},
  urldate = {2025-03-26},
  archiveprefix = {arXiv}
}

@misc{chattopadhyay_challenges_2024,
  title = {Challenges of Learning Multi-Scale Dynamics with {{AI}} Weather Models: {{Implications}} for Stability and One Solution},
  shorttitle = {Challenges of Learning Multi-Scale Dynamics with {{AI}} Weather Models},
  author = {Chattopadhyay, Ashesh and Sun, Y. Qiang and Hassanzadeh, Pedram},
  year = 2024,
  month = dec,
  number = {arXiv:2304.07029},
  eprint = {2304.07029},
  primaryclass = {physics},
  publisher = {arXiv},
  doi = {10.48550/arXiv.2304.07029},
  urldate = {2025-03-26},
  archiveprefix = {arXiv}
}

@misc{chen_analog_2023,
  title = {Analog {{Bits}}: {{Generating Discrete Data}} Using {{Diffusion Models}} with {{Self-Conditioning}}},
  shorttitle = {Analog {{Bits}}},
  author = {Chen, Ting and Zhang, Ruixiang and Hinton, Geoffrey},
  year = 2023,
  month = feb,
  number = {arXiv:2208.04202},
  eprint = {2208.04202},
  primaryclass = {cs},
  publisher = {arXiv},
  doi = {10.48550/arXiv.2208.04202},
  urldate = {2024-01-16},
  archiveprefix = {arXiv}
}

@article{chen_multivariate_2024,
  title = {Multivariate State and Parameter Estimation with Data Assimilation Applied to Sea-Ice Models Using a {{Maxwell}} Elasto-Brittle Rheology},
  author = {Chen, Yumeng and Smith, Polly and Carrassi, Alberto and Pasmans, Ivo and Bertino, Laurent and Bocquet, Marc and Finn, Tobias Sebastian and Rampal, Pierre and Dansereau, V{\'e}ronique},
  year = 2024,
  month = may,
  journal = {The Cryosphere},
  volume = {18},
  number = {5},
  pages = {2381--2406},
  publisher = {Copernicus GmbH},
  issn = {1994-0416},
  doi = {10.5194/tc-18-2381-2024},
  urldate = {2024-09-30},
  langid = {english}
}

@misc{chen_stabilized_2025,
  title = {Towards Stabilized and Efficient Diffusion Transformers through Long-Skip-Connections with Spectral Constraints},
  author = {Chen, Guanjie and Zhao, Xinyu and Zhou, Yucheng and Qu, Xiaoye and Chen, Tianlong and Cheng, Yu},
  year = 2025,
  month = jul,
  number = {arXiv:2411.17616},
  eprint = {2411.17616},
  primaryclass = {cs},
  publisher = {arXiv},
  doi = {10.48550/arXiv.2411.17616},
  urldate = {2025-08-14},
  archiveprefix = {arXiv}
}

@inproceedings{cipolla_multitask_2018,
  title = {Multi-Task {{Learning Using Uncertainty}} to {{Weigh Losses}} for {{Scene Geometry}} and {{Semantics}}},
  booktitle = {2018 {{IEEE}}/{{CVF Conference}} on {{Computer Vision}} and {{Pattern Recognition}}},
  author = {Cipolla, Roberto and Gal, Yarin and Kendall, Alex},
  year = 2018,
  month = jun,
  pages = {7482--7491},
  publisher = {IEEE},
  address = {Salt Lake City, UT, USA},
  doi = {10.1109/CVPR.2018.00781},
  urldate = {2022-05-17},
  isbn = {978-1-5386-6420-9},
  langid = {english}
}

@misc{couairon_archesweather_2024,
  title = {{{ArchesWeather}} \& {{ArchesWeatherGen}}: {{A}} Deterministic and Generative Model for Efficient {{ML}} Weather Forecasting},
  shorttitle = {{{ArchesWeather}} \& {{ArchesWeatherGen}}},
  author = {Couairon, Guillaume and Singh, Renu and Charantonis, Anastase and Lessig, Christian and Monteleoni, Claire},
  year = 2024,
  month = dec,
  number = {arXiv:2412.12971},
  eprint = {2412.12971},
  primaryclass = {cs},
  publisher = {arXiv},
  doi = {10.48550/arXiv.2412.12971},
  urldate = {2025-01-29},
  archiveprefix = {arXiv}
}

@inproceedings{crowson_scalable_2024,
  title = {Scalable {{High-Resolution Pixel-Space Image Synthesis}} with {{Hourglass Diffusion Transformers}}},
  booktitle = {Forty-First {{International Conference}} on {{Machine Learning}}},
  author = {Crowson, Katherine and Baumann, Stefan Andreas and Birch, Alex and Abraham, Tanishq Mathew and Kaplan, Daniel Z. and Shippole, Enrico},
  year = 2024,
  month = jun,
  urldate = {2024-09-13},
  langid = {english}
}

@article{dansereau_maxwell_2016,
  title = {A {{Maxwell}} Elasto-Brittle Rheology for Sea Ice Modelling},
  author = {Dansereau, V{\'e}ronique and Weiss, J{\'e}r{\^o}me and Saramito, Pierre and Lattes, Philippe},
  year = 2016,
  month = jul,
  journal = {The Cryosphere},
  volume = {10},
  number = {3},
  pages = {1339--1359},
  publisher = {Copernicus GmbH},
  issn = {1994-0424},
  doi = {10.5194/tc-10-1339-2016},
  urldate = {2021-11-16},
  langid = {english}
}

@article{durand_datadriven_2024,
  title = {Data-Driven Surrogate Modeling of High-Resolution Sea-Ice Thickness in the {{Arctic}}},
  author = {Durand, Charlotte and Finn, Tobias Sebastian and Farchi, Alban and Bocquet, Marc and Boutin, Guillaume and {\'O}lason, Einar},
  year = 2024,
  month = apr,
  journal = {The Cryosphere},
  volume = {18},
  number = {4},
  pages = {1791--1815},
  publisher = {Copernicus GmbH},
  issn = {1994-0424},
  doi = {10.5194/tc-18-1791-2024},
  urldate = {2024-04-30},
  langid = {english}
}

@article{durand_fourdimensional_2025,
  title = {Four-Dimensional Variational Data Assimilation with a Sea-Ice Thickness Emulator},
  author = {Durand, Charlotte and Finn, Tobias Sebastian and Farchi, Alban and Bocquet, Marc and Brajard, Julien and Bertino, Laurent},
  year = 2025,
  month = feb,
  journal = {EGUsphere},
  pages = {1--36},
  publisher = {Copernicus GmbH},
  doi = {10.5194/egusphere-2024-4028},
  urldate = {2025-04-03},
  langid = {english}
}

@misc{eijkelboom_variational_2024,
  title = {Variational {{Flow Matching}} for {{Graph Generation}}},
  author = {Eijkelboom, Floor and Bartosh, Grigory and Naesseth, Christian Andersson and Welling, Max and {van de Meent}, Jan-Willem},
  year = 2024,
  month = jun,
  number = {arXiv:2406.04843},
  eprint = {2406.04843},
  primaryclass = {cs, stat},
  publisher = {arXiv},
  doi = {10.48550/arXiv.2406.04843},
  urldate = {2024-09-19},
  archiveprefix = {arXiv}
}

@article{england_minimal_2025,
  title = {Minimal Arctic Sea Ice Loss in the Last 20 Years, Consistent with Internal Climate Variability},
  author = {England, M. R. and Polvani, L. M. and Screen, J. and Chan, A. C.},
  year = 2025,
  journal = {Geophysical Research Letters},
  volume = {52},
  number = {15},
  pages = {e2025GL116175},
  issn = {1944-8007},
  doi = {10.1029/2025GL116175},
  urldate = {2025-08-13},
  copyright = {\copyright{} 2025. The Author(s).},
  langid = {english}
}

@misc{falcon_pytorchlightning_2020,
  title = {{{PyTorchLightning}}: 0.7.6 Release},
  shorttitle = {{{PyTorchLightning}}/Pytorch-Lightning},
  author = {Falcon, William and Borovec, Jirka and W{\"a}lchli, Adrian and Eggert, Nic and Schock, Justus and Jordan, Jeremy and Skafte, Nicki and Ir1dXD and Bereznyuk, Vadim and Harris, Ethan and Murrell, Tullie and Yu, Peter and Pr{\ae}sius, Sebastian and Addair, Travis and Zhong, Jacob and Lipin, Dmitry and Uchida, So and Bapat, Shreyas and Schr{\"o}ter, Hendrik and Dayma, Boris and Karnachev, Alexey and Kulkarni, Akshay and Komatsu, Shunta and Martin.B and SCHIRATTI, Jean-Baptiste and Mary, Hadrien and Byrne, Donal and Eyzaguirre, Cristobal and {cinjon} and Bakhtin, Anton},
  year = 2020,
  month = may,
  doi = {10.5281/zenodo.3828935},
  urldate = {2022-08-22},
  howpublished = {Zenodo}
}

@misc{finn_diffusion_2024,
  title = {Towards Diffusion Models for Large-Scale Sea-Ice Modelling},
  author = {Finn, Tobias Sebastian and Durand, Charlotte and Farchi, Alban and Bocquet, Marc and Brajard, Julien},
  year = 2024,
  month = jun,
  number = {arXiv:2406.18417},
  eprint = {2406.18417},
  primaryclass = {physics},
  publisher = {arXiv},
  doi = {10.48550/arXiv.2406.18417},
  urldate = {2024-07-10},
  archiveprefix = {arXiv}
}

@article{finn_generative_2024,
  title = {Generative {{Diffusion}} for {{Regional Surrogate Models From Sea}}-{{Ice Simulations}}},
  author = {Finn, Tobias Sebastian and Durand, Charlotte and Farchi, Alban and Bocquet, Marc and Rampal, Pierre and Carrassi, Alberto},
  year = 2024,
  month = oct,
  journal = {Journal of Advances in Modeling Earth Systems},
  volume = {16},
  number = {10},
  pages = {e2024MS004395},
  issn = {1942-2466, 1942-2466},
  doi = {10.1029/2024MS004395},
  urldate = {2024-11-20},
  copyright = {\copyright{} 2024 The Author(s). Journal of Advances in Modeling Earth Systems published by Wiley Periodicals LLC on behalf of American Geophysical Union.},
  langid = {english}
}

@article{finn_representation_2024,
  title = {Representation Learning with Unconditional Denoising Diffusion Models for Dynamical Systems},
  author = {Finn, Tobias Sebastian and Disson, Lucas and Farchi, Alban and Bocquet, Marc and Durand, Charlotte},
  year = 2024,
  month = sep,
  journal = {Nonlinear Processes in Geophysics},
  volume = {31},
  number = {3},
  pages = {409--431},
  publisher = {Copernicus GmbH},
  issn = {1607-7946},
  doi = {10.5194/npg-31-409-2024},
  urldate = {2024-09-19},
  langid = {english}
}

@article{fishman_diffusion_2023,
  title = {Diffusion Models for Constrained Domains},
  author = {Fishman, Nic and Klarner, Leo and Bortoli, Valentin De and Mathieu, Emile and Hutchinson, Michael John},
  year = 2023,
  month = may,
  journal = {Transactions on Machine Learning Research},
  issn = {2835-8856},
  urldate = {2025-04-21},
  langid = {english}
}

@article{fortin_why_2014,
  title = {Why {{Should Ensemble Spread Match}} the {{RMSE}} of the {{Ensemble Mean}}?},
  author = {Fortin, V. and Abaza, M. and Anctil, F. and Turcotte, R.},
  year = 2014,
  month = aug,
  journal = {Journal of Hydrometeorology},
  volume = {15},
  number = {4},
  pages = {1708--1713},
  publisher = {American Meteorological Society},
  issn = {1525-7541, 1525-755X},
  doi = {10.1175/JHM-D-14-0008.1},
  urldate = {2023-02-16},
  chapter = {Journal of Hydrometeorology},
  langid = {english}
}

@article{germann_scaledependence_2002,
  title = {Scale-{{Dependence}} of the {{Predictability}} of {{Precipitation}} from {{Continental Radar Images}}. {{Part I}}: {{Description}} of the {{Methodology}}},
  shorttitle = {Scale-{{Dependence}} of the {{Predictability}} of {{Precipitation}} from {{Continental Radar Images}}. {{Part I}}},
  author = {Germann, Urs and Zawadzki, Isztar},
  year = 2002,
  month = dec,
  journal = {Monthly Weather Review},
  volume = {130},
  number = {12},
  pages = {2859--2873},
  publisher = {American Meteorological Society},
  issn = {1520-0493, 0027-0644},
  doi = {10.1175/1520-0493(2002)130<2859:SDOTPO>2.0.CO;2},
  urldate = {2024-02-10},
  chapter = {Monthly Weather Review},
  langid = {english}
}

@article{he_deep_2015,
  title = {Deep {{Residual Learning}} for {{Image Recognition}}},
  author = {He, Kaiming and Zhang, Xiangyu and Ren, Shaoqing and Sun, Jian},
  year = 2015,
  month = dec,
  journal = {arXiv:1512.03385 [cs]},
  eprint = {1512.03385},
  primaryclass = {cs},
  urldate = {2019-11-15},
  archiveprefix = {arXiv}
}

@misc{hendrycks_gaussian_2016,
  title = {Gaussian Error Linear Units (Gelus)},
  author = {Hendrycks, Dan and Gimpel, Kevin},
  year = 2016,
  number = {arXiv:1606.08415},
  eprint = {1606.08415},
  publisher = {arXiv},
  doi = {10.48550/arXiv.1606.08415},
  archiveprefix = {arXiv}
}

@misc{heo_rotary_2024,
  title = {Rotary {{Position Embedding}} for {{Vision Transformer}}},
  author = {Heo, Byeongho and Park, Song and Han, Dongyoon and Yun, Sangdoo},
  year = 2024,
  month = jul,
  number = {arXiv:2403.13298},
  eprint = {2403.13298},
  primaryclass = {cs},
  publisher = {arXiv},
  doi = {10.48550/arXiv.2403.13298},
  urldate = {2024-07-22},
  archiveprefix = {arXiv}
}

@article{hersbach_era5_2020,
  title = {The {{ERA5}} Global Reanalysis},
  author = {Hersbach, Hans and Bell, Bill and Berrisford, Paul and Hirahara, Shoji and Hor{\'a}nyi, Andr{\'a}s and Mu{\~n}oz-Sabater, Joaqu{\'i}n and Nicolas, Julien and Peubey, Carole and Radu, Raluca and Schepers, Dinand and Simmons, Adrian and Soci, Cornel and Abdalla, Saleh and Abellan, Xavier and Balsamo, Gianpaolo and Bechtold, Peter and Biavati, Gionata and Bidlot, Jean and Bonavita, Massimo and De Chiara, Giovanna and Dahlgren, Per and Dee, Dick and Diamantakis, Michail and Dragani, Rossana and Flemming, Johannes and Forbes, Richard and Fuentes, Manuel and Geer, Alan and Haimberger, Leo and Healy, Sean and Hogan, Robin J. and H{\'o}lm, El{\'i}as and Janiskov{\'a}, Marta and Keeley, Sarah and Laloyaux, Patrick and Lopez, Philippe and Lupu, Cristina and Radnoti, Gabor and De Rosnay, Patricia and Rozum, Iryna and Vamborg, Freja and Villaume, Sebastien and Th{\'e}paut, Jean-No{\"e}l},
  year = 2020,
  month = jul,
  journal = {Quarterly Journal of the Royal Meteorological Society},
  volume = {146},
  number = {730},
  pages = {1999--2049},
  issn = {0035-9009, 1477-870X},
  doi = {10.1002/qj.3803},
  urldate = {2021-05-30},
  copyright = {\copyright{} 2020 The Authors. Quarterly Journal of the Royal Meteorological Society published by John Wiley \& Sons Ltd on behalf of the Royal Meteorological Society.},
  langid = {english}
}

@article{hibler_dynamic_1979,
  title = {A {{Dynamic Thermodynamic Sea Ice Model}}},
  author = {Hibler, W. D.},
  year = 1979,
  month = jul,
  journal = {Journal of Physical Oceanography},
  volume = {9},
  number = {4},
  pages = {815--846},
  publisher = {American Meteorological Society},
  issn = {0022-3670, 1520-0485},
  doi = {10.1175/1520-0485(1979)009<0815:ADTSIM>2.0.CO;2},
  urldate = {2024-02-07},
  chapter = {Journal of Physical Oceanography},
  langid = {english}
}

@misc{hoogeboom_simple_2023,
  title = {Simple Diffusion: {{End-to-end}} Diffusion for High Resolution Images},
  shorttitle = {Simple Diffusion},
  author = {Hoogeboom, Emiel and Heek, Jonathan and Salimans, Tim},
  year = 2023,
  month = jan,
  number = {arXiv:2301.11093},
  eprint = {2301.11093},
  primaryclass = {cs, stat},
  publisher = {arXiv},
  doi = {10.48550/arXiv.2301.11093},
  urldate = {2023-11-02},
  archiveprefix = {arXiv}
}

@inproceedings{jun_distribution_2020,
  title = {Distribution {{Augmentation}} for {{Generative Modeling}}},
  booktitle = {Proceedings of the 37th {{International Conference}} on {{Machine Learning}}},
  author = {Jun, Heewoo and Child, Rewon and Chen, Mark and Schulman, John and Ramesh, Aditya and Radford, Alec and Sutskever, Ilya},
  year = 2020,
  month = nov,
  pages = {5006--5019},
  publisher = {PMLR},
  issn = {2640-3498},
  urldate = {2024-02-19},
  langid = {english}
}

@misc{karras_analyzing_2023,
  title = {Analyzing and {{Improving}} the {{Training Dynamics}} of {{Diffusion Models}}},
  author = {Karras, Tero and Aittala, Miika and Lehtinen, Jaakko and Hellsten, Janne and Aila, Timo and Laine, Samuli},
  year = 2023,
  month = dec,
  number = {arXiv:2312.02696},
  eprint = {2312.02696},
  primaryclass = {cs, stat},
  publisher = {arXiv},
  doi = {10.48550/arXiv.2312.02696},
  urldate = {2024-02-20},
  archiveprefix = {arXiv}
}

@misc{karras_elucidating_2022,
  title = {Elucidating the {{Design Space}} of {{Diffusion-Based Generative Models}}},
  author = {Karras, Tero and Aittala, Miika and Aila, Timo and Laine, Samuli},
  year = 2022,
  month = oct,
  number = {arXiv:2206.00364},
  eprint = {2206.00364},
  primaryclass = {cs, stat},
  publisher = {arXiv},
  doi = {10.48550/arXiv.2206.00364},
  urldate = {2022-10-27},
  archiveprefix = {arXiv}
}

@article{kingma_adam_2017,
  title = {Adam: {{A Method}} for {{Stochastic Optimization}}},
  shorttitle = {Adam},
  author = {Kingma, Diederik P. and Ba, Jimmy},
  year = 2017,
  month = jan,
  journal = {arXiv:1412.6980 [cs]},
  eprint = {1412.6980},
  primaryclass = {cs},
  urldate = {2020-09-26},
  archiveprefix = {arXiv}
}

@misc{kingma_understanding_2023,
  title = {Understanding {{Diffusion Objectives}} as the {{ELBO}} with {{Simple Data Augmentation}}},
  author = {Kingma, Diederik P. and Gao, Ruiqi},
  year = 2023,
  month = sep,
  number = {arXiv:2303.00848},
  eprint = {2303.00848},
  primaryclass = {cs, stat},
  publisher = {arXiv},
  doi = {10.48550/arXiv.2303.00848},
  urldate = {2023-10-05},
  archiveprefix = {arXiv}
}

@article{kochkov_neural_2024,
  title = {Neural General Circulation Models for Weather and Climate},
  author = {Kochkov, Dmitrii and Yuval, Janni and Langmore, Ian and Norgaard, Peter and Smith, Jamie and Mooers, Griffin and Kl{\"o}wer, Milan and Lottes, James and Rasp, Stephan and D{\"u}ben, Peter and Hatfield, Sam and Battaglia, Peter and {Sanchez-Gonzalez}, Alvaro and Willson, Matthew and Brenner, Michael P. and Hoyer, Stephan},
  year = 2024,
  month = aug,
  journal = {Nature},
  volume = {632},
  number = {8027},
  pages = {1060--1066},
  publisher = {Nature Publishing Group},
  issn = {0028-0836, 1476-4687},
  doi = {10.1038/s41586-024-07744-y},
  urldate = {2024-07-30},
  copyright = {2024 The Author(s)},
  langid = {english}
}

@article{kvanum_developing_2024,
  title = {Developing a Deep Learning Forecasting System for Short-Term and High-Resolution Prediction of Sea Ice Concentration},
  author = {Kvanum, Are Frode and Palerme, Cyril and M{\"u}ller, Malte and Rabault, Jean and Hughes, Nick},
  year = 2024,
  month = feb,
  journal = {EGUsphere},
  pages = {1--26},
  publisher = {Copernicus GmbH},
  doi = {10.5194/egusphere-2023-3107},
  urldate = {2024-07-30},
  langid = {english}
}

@misc{lam_graphcast_2023,
  title = {{{GraphCast}}: {{Learning}} Skillful Medium-Range Global Weather Forecasting},
  shorttitle = {{{GraphCast}}},
  author = {Lam, Remi and {Sanchez-Gonzalez}, Alvaro and Willson, Matthew and Wirnsberger, Peter and Fortunato, Meire and Alet, Ferran and Ravuri, Suman and Ewalds, Timo and {Eaton-Rosen}, Zach and Hu, Weihua and Merose, Alexander and Hoyer, Stephan and Holland, George and Vinyals, Oriol and Stott, Jacklynn and Pritzel, Alexander and Mohamed, Shakir and Battaglia, Peter},
  year = 2023,
  month = aug,
  number = {arXiv:2212.12794},
  eprint = {2212.12794},
  primaryclass = {physics},
  publisher = {arXiv},
  doi = {10.48550/arXiv.2212.12794},
  urldate = {2023-11-09},
  archiveprefix = {arXiv}
}

@article{lam_learning_2023,
  title = {Learning Skillful Medium-Range Global Weather Forecasting},
  author = {Lam, Remi and {Sanchez-Gonzalez}, Alvaro and Willson, Matthew and Wirnsberger, Peter and Fortunato, Meire and Alet, Ferran and Ravuri, Suman and Ewalds, Timo and {Eaton-Rosen}, Zach and Hu, Weihua and Merose, Alexander and Hoyer, Stephan and Holland, George and Vinyals, Oriol and Stott, Jacklynn and Pritzel, Alexander and Mohamed, Shakir and Battaglia, Peter},
  year = 2023,
  month = dec,
  journal = {Science},
  volume = {382},
  number = {6677},
  pages = {1416--1421},
  publisher = {American Association for the Advancement of Science},
  issn = {0036-8075, 1095-9203},
  doi = {10.1126/science.adi2336},
  urldate = {2023-11-15}
}

@misc{lang_aifs_2024,
  title = {{{AIFS}} - {{ECMWF}}'s Data-Driven Forecasting System},
  author = {Lang, Simon and Alexe, Mihai and Chantry, Matthew and Dramsch, Jesper and Pinault, Florian and Raoult, Baudouin and Clare, Mariana C. A. and Lessig, Christian and {Maier-Gerber}, Michael and Magnusson, Linus and Bouall{\`e}gue, Zied Ben and Nemesio, Ana Prieto and Dueben, Peter D. and Brown, Andrew and Pappenberger, Florian and Rabier, Florence},
  year = 2024,
  month = aug,
  number = {arXiv:2406.01465},
  eprint = {2406.01465},
  primaryclass = {physics},
  publisher = {arXiv},
  doi = {10.48550/arXiv.2406.01465},
  urldate = {2024-08-14},
  archiveprefix = {arXiv}
}

@article{lepparanta_review_1993,
  title = {A Review of Analytical Models of Sea-ice Growth},
  author = {Lepp{\"a}ranta, Matti},
  year = 1993,
  month = mar,
  journal = {Atmosphere-Ocean},
  volume = {31},
  number = {1},
  pages = {123--138},
  publisher = {Taylor \& Francis},
  issn = {0705-5900},
  doi = {10.1080/07055900.1993.9649465},
  urldate = {2025-08-13}
}

@misc{li_seeds_2023,
  title = {{{SEEDS}}: {{Emulation}} of {{Weather Forecast Ensembles}} with {{Diffusion Models}}},
  shorttitle = {{{SEEDS}}},
  author = {Li, Lizao and Carver, Rob and {Lopez-Gomez}, Ignacio and Sha, Fei and Anderson, John},
  year = 2023,
  month = oct,
  number = {arXiv:2306.14066},
  eprint = {2306.14066},
  primaryclass = {physics},
  publisher = {arXiv},
  doi = {10.48550/arXiv.2306.14066},
  urldate = {2023-11-30},
  archiveprefix = {arXiv}
}

@misc{lipman_flow_2023,
  title = {Flow {{Matching}} for {{Generative Modeling}}},
  author = {Lipman, Yaron and Chen, Ricky T. Q. and {Ben-Hamu}, Heli and Nickel, Maximilian and Le, Matt},
  year = 2023,
  month = feb,
  number = {arXiv:2210.02747},
  eprint = {2210.02747},
  primaryclass = {cs, stat},
  publisher = {arXiv},
  doi = {10.48550/arXiv.2210.02747},
  urldate = {2024-03-14},
  archiveprefix = {arXiv}
}

@article{liu_daily_2021,
  title = {Daily {{Prediction}} of the {{Arctic Sea Ice Concentration Using Reanalysis Data Based}} on a {{Convolutional LSTM Network}}},
  author = {Liu, Quanhong and Zhang, Ren and Wang, Yangjun and Yan, Hengqian and Hong, Mei},
  year = 2021,
  month = mar,
  journal = {Journal of Marine Science and Engineering},
  volume = {9},
  number = {3},
  pages = {330},
  publisher = {Multidisciplinary Digital Publishing Institute},
  issn = {2077-1312},
  doi = {10.3390/jmse9030330},
  urldate = {2022-10-04},
  copyright = {http://creativecommons.org/licenses/by/3.0/},
  langid = {english}
}

@misc{liu_flow_2022,
  title = {Flow {{Straight}} and {{Fast}}: {{Learning}} to {{Generate}} and {{Transfer Data}} with {{Rectified Flow}}},
  shorttitle = {Flow {{Straight}} and {{Fast}}},
  author = {Liu, Xingchao and Gong, Chengyue and Liu, Qiang},
  year = 2022,
  month = sep,
  number = {arXiv:2209.03003},
  eprint = {2209.03003},
  primaryclass = {cs},
  publisher = {arXiv},
  doi = {10.48550/arXiv.2209.03003},
  urldate = {2024-07-17},
  archiveprefix = {arXiv}
}

@misc{loshchilov_decoupled_2019,
  title = {Decoupled {{Weight Decay Regularization}}},
  author = {Loshchilov, Ilya and Hutter, Frank},
  year = 2019,
  month = jan,
  number = {arXiv:1711.05101},
  eprint = {1711.05101},
  primaryclass = {cs, math},
  publisher = {arXiv},
  doi = {10.48550/arXiv.1711.05101},
  urldate = {2024-02-19},
  archiveprefix = {arXiv}
}

@inproceedings{lou_reflected_2023,
  title = {Reflected {{Diffusion Models}}},
  booktitle = {Proceedings of the 40th {{International Conference}} on {{Machine Learning}}},
  author = {Lou, Aaron and Ermon, Stefano},
  year = 2023,
  month = jul,
  pages = {22675--22701},
  publisher = {PMLR},
  issn = {2640-3498},
  urldate = {2024-06-24},
  langid = {english}
}

@article{madec_nemo_2023,
  title = {{{NEMO Ocean Engine Reference Manual}}},
  author = {Madec, Gurvan and Bell, Mike and Blaker, Adam and Bricaud, Cl{\'e}ment and Bruciaferri, Diego and Castrillo, Miguel and Calvert, Daley and {J\'er\^omeme Chanut} and Clementi, Emanuela and Coward, Andrew and Epicoco, Italo and {\'E}th{\'e}, Christian and Ganderton, Jonas and Harle, James and Hutchinson, Katherine and Iovino, Doroteaciro and Lea, Dan and Lovato, Tomas and Martin, Matt and Martin, Nicolas and Mele, Francesca and Martins, Diana and Masson, S{\'e}bastien and Mathiot, Pierre and Mocavero, Silvia and M{\"u}ller, Simon and Nurser, A.J. George and Paronuzzi, Stella and Peltier, Mathieu and Person, Renaud and Rousset, Clement and Rynders, Stefanie and Samson, Guillaume and T{\'e}chen{\'e}, Sibylle and Vancoppenolle, Martin and Wilson, Chris},
  year = 2023,
  month = jul,
  publisher = {Zenodo},
  issn = {1288-1619},
  doi = {10.5281/zenodo.8167700},
  urldate = {2024-02-09},
  langid = {english}
}

@misc{mardani_residual_2024,
  title = {Residual Corrective Diffusion Modeling for Km-Scale Atmospheric Downscaling},
  author = {Mardani, Morteza and Brenowitz, Noah and Cohen, Yair and Pathak, Jaideep and Chen, Chieh-Yu and Liu, Cheng-Chin and Vahdat, Arash and Nabian, Mohammad Amin and Ge, Tao and Subramaniam, Akshay and Kashinath, Karthik and Kautz, Jan and Pritchard, Mike},
  year = 2024,
  month = aug,
  number = {arXiv:2309.15214},
  eprint = {2309.15214},
  publisher = {arXiv},
  doi = {10.48550/arXiv.2309.15214},
  urldate = {2024-10-22},
  archiveprefix = {arXiv}
}

@article{marsan_scale_2004,
  title = {Scale {{Dependence}} and {{Localization}} of the {{Deformation}} of {{Arctic Sea Ice}}},
  author = {Marsan, David and Stern, Harry and Lindsay, Ron and Weiss, J{\'e}r{\^o}me},
  year = 2004,
  month = oct,
  journal = {Physical Review Letters},
  volume = {93},
  number = {17},
  pages = {178501},
  publisher = {American Physical Society},
  issn = {0031-9007, 1079-7114},
  doi = {10.1103/PhysRevLett.93.178501},
  urldate = {2024-03-13}
}

@misc{meier_arctic_2023,
  title = {Arctic and Antarctic Regional Masks for Sea Ice and Related Data Products, Version 1},
  author = {Meier, Walter and Stewart, J},
  year = 2023,
  publisher = {{NASA National Snow and Ice Data Center Distributed Active Archive Center}},
  doi = {10.5067/CYW3O8ZUNIWC}
}

@misc{meier_noaa_2024,
  title = {{{NOAA}}/{{NSIDC}} Climate Data Record of Passive Microwave Sea Ice Concentration, Version 5},
  author = {Meier, Walt and Fetterer, Florence and Windnagel, Ann and Stewart, James Scott and Stafford, Trey},
  year = 2024,
  publisher = {{National Snow and Ice Data Center}},
  doi = {10.7265/RJZB-PF78}
}

@misc{meuer_latent_2024,
  title = {Latent {{Diffusion Model}} for {{Generating Ensembles}} of {{Climate Simulations}}},
  author = {Meuer, Johannes and Witte, Maximilian and Finn, Tobias Sebastian and Timmreck, Claudia and Ludwig, Thomas and Kadow, Christopher},
  year = 2024,
  month = jul,
  number = {arXiv:2407.02070},
  eprint = {2407.02070},
  primaryclass = {physics},
  publisher = {arXiv},
  doi = {10.48550/arXiv.2407.02070},
  urldate = {2024-07-10},
  archiveprefix = {arXiv}
}

@article{mohammadi-aragh_predictability_2018,
  title = {Predictability of Arctic Sea Ice on Weather Time Scales},
  author = {{Mohammadi-Aragh}, M. and Goessling, H. F. and Losch, M. and Hutter, N. and Jung, T.},
  year = 2018,
  month = apr,
  journal = {Scientific Reports},
  volume = {8},
  number = {1},
  pages = {6514},
  publisher = {Nature Publishing Group},
  issn = {2045-2322},
  doi = {10.1038/s41598-018-24660-0},
  urldate = {2025-08-19},
  copyright = {2018 The Author(s)},
  langid = {english}
}

@article{moldovan_aifs_2025,
  title = {{{AIFS}} 1.1.0: {{An}} Update to {{ECMWF}}'s Machine-Learned Weather Forecast Model {{AIFS}}},
  shorttitle = {{{AIFS}} 1.1.0},
  author = {Moldovan, Gabriel and Pinnington, Ewan and Prieto Nemesio, Ana and Lang, Simon and Ben Bouall{\`e}gue, Zied and Dramsch, Jesper and Alexe, Mihai and Santa Cruz, Mario and Hahner, Sara and Cook, Harrison and Theissen, Helen and Clare, Mariana and O'Brien, Cathal and Polster, Jan and Magnusson, Linus and Mertes, Gert and Pinault, Florian and Raoult, Baudouin and {de Rosnay}, Patricia and Forbes, Richard and Chantry, Matthew},
  year = 2025,
  month = oct,
  journal = {EGUsphere},
  pages = {1--23},
  publisher = {Copernicus GmbH},
  doi = {10.5194/egusphere-2025-4716},
  urldate = {2025-11-05},
  langid = {english}
}

@article{olason_new_2022,
  title = {A {{New Brittle Rheology}} and {{Numerical Framework}} for {{Large}}-{{Scale Sea}}-{{Ice Models}}},
  author = {{\'O}lason, Einar and Boutin, Guillaume and Korosov, Anton and Rampal, Pierre and Williams, Timothy and Kimmritz, Madlen and Dansereau, V{\'e}ronique and Samak{\'e}, Abdoulaye},
  year = 2022,
  month = aug,
  journal = {Journal of Advances in Modeling Earth Systems},
  volume = {14},
  number = {8},
  pages = {e2021MS002685},
  issn = {1942-2466, 1942-2466},
  doi = {10.1029/2021MS002685},
  urldate = {2023-12-20},
  copyright = {\copyright{} 2022 The Authors. Journal of Advances in Modeling Earth Systems published by Wiley Periodicals LLC on behalf of American Geophysical Union.},
  langid = {english}
}

@misc{osisaf_concentration_2025,
  title = {Global Sea Ice Concentration Climate Data Record 1978-2020},
  author = {{EUMETSAT Ocean {and} Sea Ice Satellite Application Facility}},
  year = 2025,
  number = {OSI-450-a1},
  doi = {10.15770/EUM_SAF_OSI_0023}
}

@misc{parthipan_defining_2024,
  title = {Defining Error Accumulation in {{ML}} Atmospheric Simulators},
  author = {Parthipan, Raghul and Anand, Mohit and Christensen, Hannah M. and Hosking, J. Scott and Wischik, Damon J.},
  year = 2024,
  month = may,
  number = {arXiv:2405.14714},
  eprint = {2405.14714},
  primaryclass = {cs},
  publisher = {arXiv},
  doi = {10.48550/arXiv.2405.14714},
  urldate = {2025-03-26},
  archiveprefix = {arXiv}
}

@incollection{paszke_pytorch_2019,
  title = {{{PyTorch}}: {{An Imperative Style}}, {{High-Performance Deep Learning Library}}},
  booktitle = {Advances in {{Neural Information Processing Systems}} 32},
  author = {Paszke, Adam and Gross, Sam and Massa, Francisco and Lerer, Adam and Bradbury, James and Chanan, Gregory and Killeen, Trevor and Lin, Zeming and Gimelshein, Natalia and Antiga, Luca and Desmaison, Alban and Kopf, Andreas and Yang, Edward and DeVito, Zachary and Raison, Martin and Tejani, Alykhan and Chilamkurthy, Sasank and Steiner, Benoit and Fang, Lu and Bai, Junjie and Chintala, Soumith},
  editor = {Wallach, H. and Larochelle, H. and Beygelzimer, A. and {Alch{\'e}-Buc}, F. and Fox, E. and Garnett, R.},
  year = 2019,
  pages = {8024--8035},
  publisher = {Curran Associates, Inc.}
}

@misc{peebles_scalable_2023,
  title = {Scalable {{Diffusion Models}} with {{Transformers}}},
  author = {Peebles, William and Xie, Saining},
  year = 2023,
  month = mar,
  number = {arXiv:2212.09748},
  eprint = {2212.09748},
  primaryclass = {cs},
  publisher = {arXiv},
  doi = {10.48550/arXiv.2212.09748},
  urldate = {2024-02-19},
  archiveprefix = {arXiv}
}

@misc{petty_icesat2_2023,
  title = {{{ICESat-2 L4}} Monthly Gridded Sea Ice Thickness, Version 3},
  author = {Petty, Alek and Kurtz, Nathan and Kwok, Ron and Markus, Thorsten and Neumann, Thomas and Keeney, Nicole},
  year = 2023,
  publisher = {{NASA National Snow and Ice Data Center Distributed Active Archive Center}}
}

@article{petty_winter_2023,
  title = {Winter Arctic Sea Ice Thickness from {{ICESat-2}}: {{Upgrades}} to Freeboard and Snow Loading Estimates and an Assessment of the First Three Winters of Data Collection},
  shorttitle = {Winter {{Arctic}} Sea Ice Thickness from {{ICESat-2}}},
  author = {Petty, Alek A. and Keeney, Nicole and Cabaj, Alex and Kushner, Paul and Bagnardi, Marco},
  year = 2023,
  month = jan,
  journal = {The Cryosphere},
  volume = {17},
  number = {1},
  pages = {127--156},
  publisher = {Copernicus GmbH},
  issn = {1994-0416},
  doi = {10.5194/tc-17-127-2023},
  urldate = {2025-09-23},
  langid = {english}
}

@article{plante_sea_2025,
  title = {A Sea Ice Deformation and Rotation Rate Dataset (2017--2023) from the Environment and Climate Change Canada Automated Sea Ice Tracking System ({{ECCC-ASITS}})},
  author = {Plante, Mathieu and Lemieux, Jean-Fran{\c c}ois and Tremblay, L. Bruno and Bouchat, Am{\'e}lie and Ringeisen, Damien and Blain, Philippe and Howell, Stephen and Brady, Mike and Komarov, Alexander S. and Duval, B{\'e}atrice and Yakuden, Lekima and Labelle, Fr{\'e}d{\'e}rique},
  year = 2025,
  month = feb,
  journal = {Earth System Science Data},
  volume = {17},
  number = {2},
  pages = {423--434},
  publisher = {Copernicus GmbH},
  issn = {1866-3508},
  doi = {10.5194/essd-17-423-2025},
  urldate = {2025-07-07},
  langid = {english}
}

@article{price_probabilistic_2025,
  title = {Probabilistic Weather Forecasting with Machine Learning},
  author = {Price, Ilan and {Sanchez-Gonzalez}, Alvaro and Alet, Ferran and Andersson, Tom R. and {El-Kadi}, Andrew and Masters, Dominic and Ewalds, Timo and Stott, Jacklynn and Mohamed, Shakir and Battaglia, Peter and Lam, Remi and Willson, Matthew},
  year = 2025,
  month = jan,
  journal = {Nature},
  volume = {637},
  number = {8044},
  pages = {84--90},
  publisher = {Nature Publishing Group},
  issn = {0028-0836, 1476-4687},
  doi = {10.1038/s41586-024-08252-9},
  urldate = {2024-12-06},
  copyright = {2024 The Author(s)},
  langid = {english}
}

@inproceedings{rahimi_random_2007,
  title = {Random Features for Large-Scale Kernel Machines},
  booktitle = {Advances in {{Neural Information Processing Systems}}},
  author = {Rahimi, Ali and Recht, Benjamin},
  year = 2007,
  volume = {20},
  publisher = {Curran Associates, Inc.},
  urldate = {2025-08-14}
}

@article{rampal_multifractal_2019,
  title = {On the Multi-Fractal Scaling Properties of Sea Ice Deformation},
  author = {Rampal, Pierre and Dansereau, V{\'e}ronique and Olason, Einar and Bouillon, Sylvain and Williams, Timothy and Korosov, Anton and Samak{\'e}, Abdoulaye},
  year = 2019,
  month = sep,
  journal = {The Cryosphere},
  volume = {13},
  number = {9},
  pages = {2457--2474},
  publisher = {Copernicus GmbH},
  issn = {1994-0424},
  doi = {10.5194/tc-13-2457-2019},
  urldate = {2023-10-15},
  langid = {english}
}

@article{rampal_nextsim_2016,
  title = {{{neXtSIM}}: A New {{Lagrangian}} Sea Ice Model},
  shorttitle = {{{neXtSIM}}},
  author = {Rampal, Pierre and Bouillon, Sylvain and {\'O}lason, Einar and Morlighem, Mathieu},
  year = 2016,
  month = may,
  journal = {The Cryosphere},
  volume = {10},
  number = {3},
  pages = {1055--1073},
  publisher = {Copernicus GmbH},
  issn = {1994-0424},
  doi = {10.5194/tc-10-1055-2016},
  urldate = {2022-06-05},
  langid = {english}
}

@article{rampal_scaling_2008,
  title = {Scaling Properties of Sea Ice Deformation from Buoy Dispersion Analysis},
  author = {Rampal, P. and Weiss, J. and Marsan, D. and Lindsay, R. and Stern, H.},
  year = 2008,
  month = mar,
  journal = {Journal of Geophysical Research: Oceans},
  volume = {113},
  number = {C3},
  pages = {2007JC004143},
  issn = {0148-0227},
  doi = {10.1029/2007JC004143},
  urldate = {2025-06-06},
  copyright = {Copyright 2008 by the American Geophysical Union.},
  langid = {english}
}

@article{ricker_weekly_2017,
  title = {A Weekly Arctic Sea-Ice Thickness Data Record from Merged {{CryoSat-2}} and {{SMOS}} Satellite Data},
  author = {Ricker, Robert and Hendricks, Stefan and Kaleschke, Lars and {Tian-Kunze}, Xiangshan and King, Jennifer and Haas, Christian},
  year = 2017,
  month = jul,
  journal = {The Cryosphere},
  volume = {11},
  number = {4},
  pages = {1607--1623},
  publisher = {Copernicus GmbH},
  issn = {1994-0416},
  doi = {10.5194/tc-11-1607-2017},
  urldate = {2025-08-13},
  langid = {english}
}

@article{ronneberger_unet_2015,
  title = {U-{{Net}}: {{Convolutional Networks}} for {{Biomedical Image Segmentation}}},
  shorttitle = {U-{{Net}}},
  author = {Ronneberger, Olaf and Fischer, Philipp and Brox, Thomas},
  year = 2015,
  month = may,
  journal = {arXiv:1505.04597 [cs]},
  eprint = {1505.04597},
  primaryclass = {cs},
  urldate = {2022-03-08},
  archiveprefix = {arXiv}
}

@article{saharia_photorealistic_2022,
  title = {Photorealistic {{Text-to-Image Diffusion Models}} with {{Deep Language Understanding}}},
  author = {Saharia, Chitwan and Chan, William and Saxena, Saurabh and Li, Lala and Whang, Jay and Denton, Emily L. and Ghasemipour, Kamyar and Gontijo Lopes, Raphael and Karagol Ayan, Burcu and Salimans, Tim and Ho, Jonathan and Fleet, David J. and Norouzi, Mohammad},
  year = 2022,
  month = dec,
  journal = {Advances in Neural Information Processing Systems},
  volume = {35},
  pages = {36479--36494},
  urldate = {2023-11-16},
  langid = {english}
}

@article{samake_parallel_2017,
  title = {Parallel Implementation of a {{Lagrangian-based}} Model on an Adaptive Mesh in {{C}}++: {{Application}} to Sea-Ice},
  shorttitle = {Parallel Implementation of a {{Lagrangian-based}} Model on an Adaptive Mesh in {{C}}++},
  author = {Samak{\'e}, Abdoulaye and Rampal, Pierre and Bouillon, Sylvain and {\'O}lason, Einar},
  year = 2017,
  month = dec,
  journal = {Journal of Computational Physics},
  volume = {350},
  pages = {84--96},
  issn = {00219991},
  doi = {10.1016/j.jcp.2017.08.055},
  urldate = {2025-03-26}
}

@misc{shazeer_glu_2020,
  title = {{{GLU}} Variants Improve Transformer},
  author = {Shazeer, Noam},
  year = 2020,
  month = feb,
  number = {arXiv:2002.05202},
  eprint = {2002.05202},
  primaryclass = {cs, stat},
  publisher = {arXiv},
  doi = {10.48550/arXiv.2002.05202},
  urldate = {2024-06-24},
  archiveprefix = {arXiv}
}

@misc{shi_realtime_2016,
  title = {Real-Time Single Image and Video Super-Resolution Using an Efficient Sub-Pixel Convolutional Neural Network},
  author = {Shi, Wenzhe and Caballero, Jose and Husz{\'a}r, Ferenc and Totz, Johannes and Aitken, Andrew P. and Bishop, Rob and Rueckert, Daniel and Wang, Zehan},
  year = 2016,
  month = sep,
  number = {arXiv:1609.05158},
  eprint = {1609.05158},
  primaryclass = {cs},
  publisher = {arXiv},
  doi = {10.48550/arXiv.1609.05158},
  urldate = {2025-08-14},
  archiveprefix = {arXiv}
}

@article{sohl-dickstein_deep_2015,
  title = {Deep {{Unsupervised Learning}} Using {{Nonequilibrium Thermodynamics}}},
  author = {{Sohl-Dickstein}, Jascha and Weiss, Eric A. and Maheswaranathan, Niru and Ganguli, Surya},
  year = 2015,
  month = nov,
  journal = {arXiv:1503.03585 [cond-mat, q-bio, stat]},
  eprint = {1503.03585},
  primaryclass = {cond-mat, q-bio, stat},
  urldate = {2022-02-24},
  archiveprefix = {arXiv}
}

@misc{song_scorebased_2021,
  title = {Score-{{Based Generative Modeling}} through {{Stochastic Differential Equations}}},
  author = {Song, Yang and {Sohl-Dickstein}, Jascha and Kingma, Diederik P. and Kumar, Abhishek and Ermon, Stefano and Poole, Ben},
  year = 2021,
  month = feb,
  number = {arXiv:2011.13456},
  eprint = {2011.13456},
  primaryclass = {cs, stat},
  publisher = {arXiv},
  doi = {10.48550/arXiv.2011.13456},
  urldate = {2023-05-03},
  archiveprefix = {arXiv}
}

@article{su_roformer_2024,
  title = {{{RoFormer}}: {{Enhanced}} Transformer with {{Rotary Position Embedding}}},
  shorttitle = {{{RoFormer}}},
  author = {Su, Jianlin and Ahmed, Murtadha and Lu, Yu and Pan, Shengfeng and Bo, Wen and Liu, Yunfeng},
  year = 2024,
  month = feb,
  journal = {Neurocomputing},
  volume = {568},
  pages = {127063},
  issn = {09252312},
  doi = {10.1016/j.neucom.2023.127063},
  urldate = {2024-07-04}
}

@article{sumata_regime_2023,
  title = {Regime Shift in {{Arctic Ocean}} Sea Ice Thickness},
  author = {Sumata, Hiroshi and De Steur, Laura and Divine, Dmitry V. and Granskog, Mats A. and Gerland, Sebastian},
  year = 2023,
  month = mar,
  journal = {Nature},
  volume = {615},
  number = {7952},
  pages = {443--449},
  publisher = {Nature Publishing Group},
  issn = {0028-0836, 1476-4687},
  doi = {10.1038/s41586-022-05686-x},
  urldate = {2025-03-26},
  copyright = {2023 The Author(s)},
  langid = {english}
}

@article{tobin_estimation_1958,
  title = {Estimation of {{Relationships}} for {{Limited Dependent Variables}}},
  author = {Tobin, James},
  year = 1958,
  month = jan,
  journal = {Econometrica},
  volume = {26},
  number = {1},
  eprint = {1907382},
  eprinttype = {jstor},
  pages = {24},
  publisher = {[Wiley, Econometric Society]},
  issn = {00129682},
  doi = {10.2307/1907382},
  urldate = {2023-10-12}
}

@misc{tschudi_easegrid_2019,
  title = {{{EASE-grid}} Sea Ice Age, Version 4},
  author = {Tschudi, Mark and Meier, Walter and Stewart, J and Fowler, Charles and Maslanik, Jim},
  year = 2019,
  publisher = {{NASA National Snow and Ice Data Center Distributed Active Archive Center}},
  doi = {10.5067/UTAV7490FEPB}
}

@techreport{vanrossum_python_1995,
  title = {Python Tutorial, {{Technical Report CS-R9526}}},
  author = {Van Rossum, Guido},
  year = 1995,
  month = may,
  address = {Amsterdam},
  institution = {Centrum voor Wiskunde en Informatica (CWI)}
}

@article{vaswani_attention_2017,
  ids = {vaswani_attention_2017-1},
  title = {Attention {{Is All You Need}}},
  author = {Vaswani, Ashish and Shazeer, Noam and Parmar, Niki and Uszkoreit, Jakob and Jones, Llion and Gomez, Aidan N. and Kaiser, Lukasz and Polosukhin, Illia},
  year = 2017,
  month = dec,
  journal = {arXiv:1706.03762 [cs]},
  eprint = {1706.03762},
  primaryclass = {cs},
  urldate = {2019-11-20},
  archiveprefix = {arXiv}
}

@inproceedings{wang_patch_2024,
  title = {Patch Diffusion: {{Faster}} and More Data-Efficient Training of Diffusion Models},
  shorttitle = {Patch Diffusion},
  booktitle = {Proceedings of the 37th {{International Conference}} on {{Neural Information Processing Systems}}},
  author = {Wang, Zhendong and Jiang, Yifan and Zheng, Huangjie and Wang, Peihao and He, Pengcheng and Wang, Zhangyang and Chen, Weizhu and Zhou, Mingyuan},
  year = 2024,
  month = may,
  series = {{{NIPS}} '23},
  pages = {72137--72154},
  publisher = {Curran Associates Inc.},
  address = {Red Hook, NY, USA},
  urldate = {2024-12-11}
}

@misc{xu_icediff_2024,
  title = {{{IceDiff}}: {{High Resolution}} and {{High-Quality Sea Ice Forecasting}} with {{Generative Diffusion Prior}}},
  shorttitle = {{{IceDiff}}},
  author = {Xu, Jingyi and Tu, Siwei and Yang, Weidong and Li, Shuhao and Liu, Keyi and Luo, Yeqi and Ma, Lipeng and Fei, Ben and Bai, Lei},
  year = 2024,
  month = oct,
  number = {arXiv:2410.09111},
  eprint = {2410.09111},
  primaryclass = {physics},
  publisher = {arXiv},
  doi = {10.48550/arXiv.2410.09111},
  urldate = {2024-12-10},
  archiveprefix = {arXiv}
}

@misc{yadan_hydra_2019,
  title = {Hydra - {{A}} Framework for Elegantly Configuring Complex Applications},
  author = {Yadan, Omry},
  year = 2019,
  howpublished = {Github}
}

@inproceedings{zhang_diffcollage_2023,
  title = {{{DiffCollage}}: {{Parallel Generation}} of {{Large Content}} with {{Diffusion Models}}},
  shorttitle = {{{DiffCollage}}},
  booktitle = {2023 {{IEEE}}/{{CVF Conference}} on {{Computer Vision}} and {{Pattern Recognition}} ({{CVPR}})},
  author = {Zhang, Qinsheng and Song, Jiaming and Huang, Xun and Chen, Yongxin and Liu, Ming-Yu},
  year = 2023,
  month = jun,
  pages = {10188--10198},
  issn = {2575-7075},
  doi = {10.1109/CVPR52729.2023.00982},
  urldate = {2024-12-11}
}

@article{zhang_modeling_2003,
  title = {Modeling Global Sea Ice with a Thickness and Enthalpy Distribution Model in Generalized Curvilinear Coordinates},
  author = {Zhang, Jinlun and Rothrock, D. A.},
  year = 2003,
  month = may,
  journal = {Monthly Weather Review},
  volume = {131},
  number = {5},
  pages = {845--861},
  publisher = {American Meteorological Society},
  issn = {1520-0493, 0027-0644},
  doi = {10.1175/1520-0493(2003)131<0845:MGSIWA>2.0.CO;2},
  urldate = {2025-08-13},
  chapter = {Monthly Weather Review},
  langid = {english}
}

@misc{zhang_root_2019,
  title = {Root {{Mean Square Layer Normalization}}},
  author = {Zhang, Biao and Sennrich, Rico},
  year = 2019,
  month = oct,
  number = {arXiv:1910.07467},
  eprint = {1910.07467},
  primaryclass = {cs, stat},
  publisher = {arXiv},
  doi = {10.48550/arXiv.1910.07467},
  urldate = {2024-06-24},
  archiveprefix = {arXiv}
}
\bibliographystyle{sciencemag}

%
%
%
%
%
%


\section*{Acknowledgments}
We would like to thank Lorenzo Zampieri, Joffrey Dumont Le Brazidec, and Steffen Tietsche (ECMWF), through discussing they helped to focus this study.
Additionally, we acknowledge Guillaume Boutin (NERSC) for providing the training dataset and other members from the SASIP project which gave helpful comments along the way.
CEREA is a member of the Institut Pierre-Simon Laplace (IPSL).

\paragraph*{Funding:}
This study is a contribution to the SASIP project funded under Grant no. \textit{G-24-66154} by Schmidt Science--a philanthropic initiative that seeks to improve societal outcomes through the development of emerging science and technologies.
TSF, MB, CD, and AF additionally received financial support from INSU/CNRS by the project DeepGeneSIS (PNTS).
This work was granted access to the HPC resources of IDRIS under the allocations 2021-AD011013069, 2022-AD011013069R1, 2023-AD011013069R2, and 2024-AD011013069R3 made by GENCI.

\paragraph*{Author contributions:}
Pierre Rampal (PR), Alberto Carrassi (AC), and Marc Bocquet (MB) acquired the project.
Tobias Sebastian Finn (TSF), MB, and Alban Farchi (AF) initialised the scientific questions based on previous work of Charlotte Durand (CD) and Flavia Porro (FP).
TSF, MB, and PR refined these scientific questions and prepared an analysis strategy.
TSF and CD organised the data.
TSF created the codebase and performed the experiments.
TSF, MB, and PR analysed and discussed the results.
TSF wrote and revised the paper, with MB, PR, CD, FP, AF, and AC reviewing.

\paragraph*{Competing interests:}
There are no competing interests to declare.

\paragraph*{Data and materials availability:}

The code for GenSIM and its training is available at \url{https://github.com/cerea-daml/gensim}, while the trained model checkpoints can be found at \url{https://huggingface.co/tobifinn/gensim}.
Scripts to produce the training data are included in the code.
The sea-ice model data used for training based on work within the SASIP project \cite{boutin_arctic_2023}.
The atmospheric forcing data is based on the ERA5 reanalysis \cite{hersbach_era5_2020} and available from the Copernicus Climate Change Service (C3S) Climate Data Store (CDS).
This work contains data modified from Copernicus Climate Change Service information.
Neither the European Commission nor ECMWF is responsible for any use that may be made of the Copernicus information or data it contains. 


\subsection*{Supplementary materials}
Materials and Methods\\
Supplementary Text\\
Figures S1 to S11\\
References \textit{(52-\arabic{enumiv})}\\ 


\newpage


\renewcommand{\thefigure}{S\arabic{figure}}
\renewcommand{\thetable}{S\arabic{table}}
\renewcommand{\theequation}{S\arabic{equation}}
\renewcommand{\thepage}{S\arabic{page}}
\setcounter{figure}{0}
\setcounter{table}{0}
\setcounter{equation}{0}
\setcounter{page}{1} 


\begin{center}
\section*{Supplementary Materials for\\ \scititle}

Tobias~Sebastian~Finn$^{\ast}$,
Marc~Bocquet,
Pierre~Rampal\\
Charlotte~Durand,
Flavia~Porro,
Alban~Farchi,
Alberto~Carrassi\\
\small$^\ast$Corresponding author. Email: tobias.finn@enpc.fr
\end{center}

\subsubsection*{This PDF file includes:}
Materials and Methods\\
Supplementary Text\\
Figures S1 to S11\\

\newpage


\subsection*{Materials and Methods}

\subsubsection*{Data used for GenSIM}

The model is trained and compared to a 24-year-long simulation (1995-2018) \cite{boutin_arctic_2023} produced by neXtSIM \cite{rampal_nextsim_2016}, an advanced sea-ice model employing Lagrangian finite elements and a brittle Bingham-Maxwell rheology \cite{olason_new_2022} (data available at \url{https://github.com/sasip-climate/catalog-shared-data-SASIP/blob/main/outputs/OPA-neXtSIM_CREG025-ILBOXE140.md}, last access: 18th August 2025).
For this simulation, neXtSIM has been coupled to OPA, the ocean component of the NEMO modelling framework \cite{madec_nemo_2023} and has been forced with the ERA5 reanalysis \cite{hersbach_era5_2020} for the atmosphere.
Although neXtSIM-OPA is initialised from a climatology, we use all the data from the start as our goal is to learn a surrogate that emulates the original system as close as possible.
The curvilinear mesh of our model corresponds to the Eulerian Arakawa-C mesh of OPA, to which the model output of neXtSIM has been interpolated.
GenSIM predicts sea-ice thickness, concentration, damage, $u$- and $v$-drift, and the snow-on-ice thickness.
All variables but the damage are averaged in a six hour window, while the damage is an instantaneous state.

In addition to the initial conditions, GenSIM takes external atmospheric forcings from the ERA5 reanalysis dataset \cite{hersbach_era5_2020}.
Originally on a latitude-longitude mesh with a $1/4^\circ$ resolution, the dataset is interpolated to our curvilinear mesh by nearest neighbour.
The forcings consist of the 2-metre temperature, the 2-metre specific humidity, and the 10-metre equatorial and meridional wind.
The specific humidity is estimated as product of the 2-metre temperature, the 2-metre dew point, and the surface pressure.

To obtain neural network inputs that contain a memory about the past development without having access to the ocean, we construct temperature degree day features from the 2-metre temperature \cite{lepparanta_review_1993, braithwaite_positive_1995}.
Using a temperature threshold of $T_{\textrm{thres}} = 271.35\,\text{K}$ (freezing point at an ocean salinity of about $32\,\text{PSU}$), we construct the positive, $\mathrm{PDD}$, and negative degree days, $\mathrm{NDD}$, by
\begin{align*}
    \mathrm{PDD}(T_{\mathrm{2m}}, \Delta t) &= \int^{t}_{t-\Delta t} \mathrm{maximum(T_{\mathrm{2m}}(s)-T_{\textrm{thres}}, 0)} {\rm d}s\\
    \text{and } \mathrm{NDD}(T_{\mathrm{2m}}, \Delta t) &= \int^{t}_{t-\Delta t} \mathrm{minimum(T_{\mathrm{2m}}(s)-T_{\textrm{thres}}, 0)} {\rm d}s,
\end{align*}
respectively.
To extract seasonal and yearly features, we use a six-hourly integration resolution with two windows into the past, $\Delta t=30\,\text{days}$ and $\Delta t = 366\,\text{days}$.
This results in four additional forcing features per forecast time, leading to 12 forcing inputs in total (four forcings for two time steps and four degree day features).

\subsubsection*{Evaluation data}

We evaluate our long-term simulation with the sea-ice--ocean reanalysis product from the Pan-Arctic Ice-Ocean Modeling and Assimilation System (PIOMAS) \cite{zhang_modeling_2003} (data available at \url{https://psc.apl.uw.edu/research/projects/arctic-sea-ice-volume-anomaly}, last access: 18th August 2025).
The dataset used here is on a one-degree latitudinal-longitudinal mesh and consists of monthly averaged sea-ice thickness and concentration values.

The sea-ice volume is additionally evaluated with the CS2/SMOS product \cite{ricker_weekly_2017} (Version 2.6, data available at \url{https://spaces.awi.de/spaces/CS2SMOS/overview}, last access: 18th August 2025), which merges satellite observations from the CryoSat2 and Soil Moisture and Ocean Salinity (SMOS) missions.
As CS2/SMOS infers solely the sea-ice thickness, it is complemented by sea-ice concentration observations of the EUMETSAT's climate data record from the OSISAF project \cite{osisaf_concentration_2025}, which merges satellite measurements from SSMR, SSM/I, and SSMIS.
Both datasets are on the EASE mesh with $25\,\text{km}$ resolution (Identifier: OSI-450-a1 and OSI-430-a, Version 3, data available at \url{https://osi-saf.eumetsat.int/products/sea-ice-products}, last access: 5th November 2025).

Also, we use the thickness satellite product of IceSAT2 \cite{petty_winter_2023, petty_icesat2_2023} (Version 3, data available at \url{https://nsidc.org/data/is2sitmogr4/versions/3}, last access: 5th November 2025) for sea-ice volume validation.
This thickness data is complemented by the sea-ice concentration from the NOAA/NSIDC climate data record \cite{meier_noaa_2024} (Version 5, data available at \url{https://nsidc.org/data/g02202/versions/5}, last access: 5th November 2025).

\subsubsection*{Surrogate modelling with censored flow matching}\label{sec:short_math}

In the following, we expose the most important equations to understand how GenSIM is trained.
For a more rigorous treatment and a mathematical derivation of the equations, we refer to Supplementary Text \ref{sec:app_mathematics}.

For a forecast window of 12~hours, based on the current state $\mathbf{x}_{t}$ and the forcings in the same window $\mathbf{F}_{t:t+12\,\mathrm{h}}$, the surrogate model predicts the sea-ice state $\mathbf{x}_{t+12\,\mathrm{h}}$--consisting of sea-ice thickness, concentration, damage, the drift in $x$- and $y$-direction, and the snow-on-ice thickness.
Instead of predicting the future state directly, the model outputs predicted tendencies $\widehat{\mathbf{z}}_{1}$, which are scaled by the per-variable climatological standard deviation of the tendencies,  $\boldsymbol{\sigma}_{\textrm{clim}}$, such that we can recover the prediction by
\begin{equation}
    \widehat{\mathbf{x}}_{t+12\,\text{h}} = \mathbf{x}_{t} + \boldsymbol{\sigma}_{\textrm{clim}} \odot\widehat{\mathbf{z}}_{1}.
\end{equation}
In our autoregressive framework, the subsequent prediction $\widehat{\mathbf{x}}_{t+24\,\text{h}}$ is performed with the previous prediction and updated forcings $\mathbf{F}_{t+12\,\mathrm{h}:t+24\,\mathrm{h}}$ as new inputs.
This way, we can extend the prediction as long as possible and desired \cite{lam_learning_2023, price_probabilistic_2025}. 

For training, we have access to a dataset with data triplets $(\mathbf{x}_{t}, \mathbf{x}_{t+12\,\mathrm{h}}, \mathbf{F}_{t:t+12\,\mathrm{h}}) \sim \mathcal{D}$.
From these data triplets, we can construct the targetted tendencies, $\mathbf{z}_{1} = (\mathbf{x}_{t+12\,\mathrm{h}}-\mathbf{x}_{t})/\boldsymbol{\sigma}_{\textrm{clim}}$.

For GenSIM, we instantiate a generative flow to produce stochastic predictions \cite{lipman_flow_2023, couairon_archesweather_2024}.
The generative flow is defined by the integration of an ordinary differential equation (ODE) in pseudo time $s \in [0, 1]$,
\begin{equation}
    \frac{\textrm{d}\mathbf{z}_{s}}{\textrm{d}s} = \mathbf{u}\label{eq:methods_flow_ode}
\end{equation}
with its generative velocity $\mathbf{u}$ and the initial noise sample at pseudo time $s=0$,
\begin{equation}
    \mathbf{z}_{0} \sim \mathcal{N}(\boldsymbol{0}, \mathbf{I}).
\end{equation}
Note that the subscript $s$ describes the dependency on the pseudo time, whereas $t$ is the real time.

During training, when we know the targetted tendencies, we construct a linear interpolant \cite{albergo_stochastic_2023} connecting the known tendencies to an initial noise sample,
\begin{equation}
    \mathbf{z}_{s} = s \mathbf{z}_{1} + (1-s)\mathbf{z}_{0}.
\end{equation}
The analytical velocity, as the pseudo-time derivative of the linear interpolant, is constant in time and reads
\begin{equation}
    \mathbf{u} = \mathbf{z} - \mathbf{z}_{0}.
\end{equation}
However, this velocity is only known during training, and we use a trained neural network as approximation for predictions such that ${\mathbf{u} \approx \widehat{\mathbf{v}}_{\boldsymbol{\theta}}(\mathbf{z}_{s}, \mathbf{x}_{t}, \mathbf{F}_{t:t+12\,\mathrm{h}}, s)}$ should hold.

Based on variational flow matching \cite{eijkelboom_variational_2024}, the commonly used loss function is reparametrised to account for an appropriate bounding of the sea-ice states with $\mathbf{x}^{L}$ as lower bound and $\mathbf{x}^{U}$ as upper bound.
For this reparametrisation, we introduce a new per-variable standard deviation $\boldsymbol{\sigma}$ which specifies the expected uncertainty of the velocity approximation and is used as learnable weighting \cite{karras_analyzing_2023}.
As the true flow is a linear interpolant, we can make a mean-field approximation, factorising the loss function over the $n_{\mathrm{vars}}=6$ variables and $n_{\mathrm{grid}}$ unmasked grid points,
\begin{equation}
    \mathcal{L}_{\mathrm{GenSIM}}(\boldsymbol{\theta}, \mathbf{s}) = \mathbb{E}_{\tau, \mathbf{z}_{0}, \mathcal{D}}\left[ \sum^{n_{\mathrm{vars}}}_{k=1}\sum^{n_{\mathrm{grid}}}_{l=1} J(\boldsymbol{\theta}, \boldsymbol{\sigma}, k, l)\right].\label{eq:methods_cfm_loss}
\end{equation}
Using a censored Gaussian as variational distribution \cite{finn_diffusion_2024}, we obtain as component cost function,
\begin{align}
    J(\boldsymbol{\theta}, \mathbf{s}, k, l) =~&\mathrm{I}(x^{L}_{k} < x_{t+12\,\text{h}, k, l} < x^{U}_{k})\left[ \frac{1}{2\cdot(\sigma_k)^2}(u_{k, l}-\widehat{v}_{\boldsymbol{\theta}, k, l})^2 + \log(\sigma_{k})\right]\label{eq:methods_cfm_loss_2}\\
    &-\mathrm{I}(x_{t+12\,\text{h}, k, l} = x^{L}_{k})\left[ \log\Phi\Big(\frac{u_{k, l}-\widehat{v}_{\boldsymbol{\theta}, k, l}}{\sigma_k}\Big)\right]\notag\\
    &-\mathrm{I}(x_{t+12\,\text{h}, k, l} = x^{U}_{k})\left[ \log\Phi\Big(\frac{\widehat{v}_{\boldsymbol{\theta}, k, l}-u_{k, l}}{\sigma_k}\Big)\right]\notag ,
\end{align}
with $\Phi$ as the normal Gaussian cumulative distribution function (CDF), $\mathrm{I}(\cdot)$ as indicator function, and $(\cdot)_{k, l}$ as scalar that is indexed for the $k$-th variable and the $l$-th grid point.
Note that the case for the cost function is changed depending on the true state at time $t+12\,\text{h}$, which differs from the common practise of using activation functions for the bounds during training \cite{durand_fourdimensional_2025, moldovan_aifs_2025}.
For variables that are unbounded like the drift, the lower and upper bound vanish and only the weighted mean-squared error is left as cost function.
When the true state is between the physical bounds, the cost function is the commonly used one for flow matching \cite{lipman_flow_2023, liu_flow_2022}, where the prediction target is the true velocity.
When the true state is exactly on the lower or upper bound, the prediction is pushed towards an increase in the probability that the bounds are exceeded in negative or positive direction, respectively.
Hence, this cost function effectively modifies the role of the predicted velocity: it regresses the true velocity and classifies if the lower or upper bound is reached.

The censored Gaussian distribution can be derived from applying a static thresholding function \cite{saharia_photorealistic_2022} to the prediction that follows a Gaussian distribution and, hence, properly accounts for physical bounding in that predictions.
To apply the neural network's output as velocity approximation, we have to project the states to $\tau = 1$ and into physical space and apply the thresholding,
\begin{align}
    \widehat{\mathbf{x}}_{t+12\,\text{h} \mid s} &= \mathbf{x}_{t} + \boldsymbol{\sigma}_{\mathrm{clim}}\odot (\mathbf{z}_{s} + (1-s) \widehat{\mathbf{v}}_{\boldsymbol{\theta}})\label{eq:methods_physical_from_latent}\\ 
    \widehat{\mathbf{x}}^{\textrm{thres}}_{t+12\,\text{h} \mid s} &= \begin{cases}
        \mathbf{x}^{L} & \text{if}~~~\widehat{\mathbf{x}}_{t+12\,\text{h} \mid s} < \mathbf{x}^{L}  \\
        \mathbf{x}^{U} & \text{if}~~~\widehat{\mathbf{x}}_{t+12\,\text{h} \mid s} > \mathbf{x}^{U}  \\
        \widehat{\mathbf{x}}_{t+12\,\text{h} \mid s} &\text{otherwise}
    \end{cases}\label{eq:methods_thresholding}
\end{align}
with $\mathbf{x}^{L}$ and $\mathbf{x}^{U}$ being the lower and upper bound, respectively.
Based on this thresholded state, we recover the thresholded velocity by reversing~Equation~\ref{eq:methods_physical_from_latent}.
This thresholded velocity can be used in any integration scheme to solve the ODE that defines the generative flow,~Equation~\ref{eq:methods_flow_ode}.
Effectively, we treat the neural network's output as median approximation for the velocity, which is thresholded if the bounds are exceeded.
This way, we can generate data that lies exactly on possible bounds, while being produced by a deterministic velocity approximation.

\subsubsection*{Experimental details}

GenSIM is based on a transformer architecture with $28.1\times10^6$ parameters, as concisely described in Section \ref{sec:gensim} and further elucidated in Appendix \ref{sec:app_arch_trans}.
The training scheme is derived in \ref{sec:short_math} and the loss function for training is~Equation~\ref{eq:methods_cfm_loss}.
The training dataset is augmented by distribution augmentation \cite{jun_distribution_2020, finn_generative_2024}: the data is randomly flipped in horizontal and vertical direction, and a -$90^{\circ}$ rotation is randomly activated.
The labels, describing which augmentation is activated, are added to the input of the neural networks.

GenSIM is trained with pseudo time steps drawn from a uniform distribution, $s \sim U(0, 1)$, while the weighting of the loss function is learned alongside the model \cite{karras_elucidating_2022}, effectively reducing the need to tune the training of the generative flow.
The samples are randomly averaged using window sizes of $(1, 2, 4, 8)$ drawn with probabilities $(0.5\overline{33}, 0.2\overline{66}, 0.1\overline{33}, 0.0\overline{66})$, and subsequently cropped into patches with $80\times80$ grid points, while ensuring that each patch contains at least $10\,\%$ unmasked grid points.
For the estimation of the loss function, we remove the $8$-pixel border around each patch to obtain a prediction size of $64\times64$.
In total, we randomly extract eight patches per sample at various averaging scales for GenSIM's training.

GenSIM is trained with mini-batches of $32$ samples and the AdamW optimiser \cite{kingma_adam_2017, loshchilov_decoupled_2019} with $\beta_{1}=0.9$ and $\beta_{2}=0.99$ and a weight decay with $\lambda=10^{-3}$.
The learning rate is linearly increased from $\gamma_{\mathrm{min}} = 10^{-6}$ to $\gamma_{\mathrm{max}} = 5 \times 10^{-4}$ within the first $5000$ iterations.
Afterwards, the learning rate is decayed by a cosine until the end of the training: $10^6$ iterations for GenSIM.
As GenSIM is robust against overfitting, we retain the model at the end of the training.
We track the exponential moving average (EMA) of the model's weights with a decay rate of $\alpha=0.9999$.
We train the models on data for $1995$--$2014$, giving $913$ iterations per training epoch, while we validate on $2015$.
The testing period $2016$--$2018$ is only revealed after the training has been finished and used to produce the results in this study.

We train on eight Nvidia A100 with $80\,\text{GB}$ memory (per-device batch size $4$), on which the training needs around $2$~days.
The models are implemented with \textit{bfloat16} mixed-precision in Python \cite{vanrossum_python_1995} based on PyTorch \cite{paszke_pytorch_2019}, PyTorch lightning \cite{falcon_pytorchlightning_2020}, and Hydra \cite{yadan_hydra_2019}.
The code for the model, the training, and to produce the training data is available at \url{https://github.com/cerea-daml/gensim}, while the trained model checkpoints are available at \url{https://huggingface.co/tobifinn/gensim}.

For inference, we use two Nvidia RTX A6000 with $48\,\text{GB}$ memory.
As sampler for GenSIM, we use a second-order Heun scheme \cite{karras_analyzing_2023} with $20$ steps and a sigmoid-like pseudo-time scheduler as visualised in Fig. \ref{fig:app_weighting}b.
Each forecast with GenSIM hence needs $39$ neural network evaluations, taking around $4.5\,\text{s}$ on an Nvidia RTX A6000.

\subsubsection*{Sea-ice volume estimation}

We normalise all sea-ice volume datasets into a common format: we use the Cartesian EASE-$25$-km mesh from OSISAF as common mesh on which the volumes are estimated.
We interpolate the sea-ice thickness and concentration of each dataset with bilinear interpolation to that mesh.

Additionally, we create a common sea-ice mask: for each original dataset, we map its land-sea mask onto the common mesh.
In addition to these land-sea masks, we interpolate the climatological EASE-$12.5$-km mask (available at \url{ftp://osisaf.met.no/reprocessed/ice/oceanmasks/}, last access: 18th August, 2025) to the common mesh, retaining only cells where sea ice was observed in the climatology.

The PIOMAS dataset has a monthly temporal resolution.
Hence, we average all sea-ice thicknesses and concentrations to the monthly resolution, before we estimate their product.
The sea-ice volume is then the product between sea-ice thickness, sea-ice concentration and cell area ($25\,\text{km}\times25\,\text{km}$).

The change in the sea-ice volume, $\mathrm{vol}_{t}$, is presumably driven by dynamics (advection) and other sources, $S_{t}$, among which are growing and melting because of thermodynamics,
\begin{equation}
    \frac{\partial \mathrm{vol}_{t}}{\partial t} + \nabla\cdot(\mathbf{u}_{t}\,\mathrm{vol}_{t}) = S_{t}.
\end{equation}
To estimate the sea-ice volume change, we use finite differences, $\frac{\partial \mathrm{vol}_{t}}{\partial t} \approx (\mathrm{vol}_{t+12\,\mathrm{h}}-\mathrm{vol}_{t})/12\,\mathrm{h}$.
Working on the Arakawa-C mesh, we further approximate the dynamical part with a second‑order centred finite‑volume advection scheme based on the volume and drift for time $t$ and assume constant dynamics until time $t+12\mathrm{h}$.
The source term $S_{t}$ is then given as difference between the true volume change and the change explained by the approximated dynamics.







\subsection*{Supplementary Text}

\subsubsection*{Short-term prediction quality}\label{app:short_term}

We test GenSIM's short-term forecasting qualities in a perfect model experiment for 2016--2018, where the simulation of neXtSIM-OPA is our truth, against which we evaluate.
Initialised from the simulation and driven by ERA5 forcings, we start a 10-day forecast once a week, effectively performing $153$ initialisations.
Running an ensemble of $16$ members from the same initial conditions and forcings, we compare the performance against persistence forecasts and against a deterministic deep learning model.

We use a U-ViT architecture \cite{hoogeboom_simple_2023, finn_generative_2024, finn_diffusion_2024} for the deterministic model, as the transformer scales quadratically with the number of grid points/tokens: a convolutional U-Net coarse-grains the pan-Arctic data, while we use a transformer in the bottleneck with almost the same configuration as for GenSIM.
Due to the additional convolutional blocks, the deterministic model has $32.2\times10^6$ parameters, slightly more than GenSIM that has $28.1\times10^6$ parameters.
To counteract the increased tendency to overfit by the deterministic model, we use the model that performs the best over the validation dataset, which is reached after around $1.5\times 10^5$ iterations.
All other experimental details but the maximum number of iterations ($5\times 10^5$ iterations, important for the learning rate decay) are the same as for GenSIM.

For a 12-hour lead time, the deterministic model predicts normalised tendencies, which are scaled by the per-variable climatological standard deviation.
The loss function is then given as the mean-squared error between the predicted scaled tendencies and the true scaled tendencies.
Each variable is effectively weighted by the inverse of its climatological variations as similarly used for other multivariate deterministic surrogate models \cite{lam_learning_2023, finn_generative_2024}.
During the autoregressive prediction, we enforce the physical bounds such as non-negativity of sea-ice thickness by clipping \cite{durand_datadriven_2024}.
Hence, this architecture can be seen as a multivariate and larger version of previously used deterministic surrogate models for sea-ice predictions \cite{durand_datadriven_2024, durand_fourdimensional_2025}.

We evaluate the root-mean-squared error (RMSE) between the forecasts and the truth over all ocean grid points that are covered by sea ice in the neXtSIM-OPA simulation (concentration $>0.15$).
To estimate a normalised RMSE, we normalise the forecast errors by the standard deviation of the climatology.
The spread-skill ratio is estimated as ratio between the root-mean-squared standard deviation from GenSIM's ensemble members and the RMSE of the ensemble mean with a correction factor of $\sqrt{17/16}$ \cite{fortin_why_2014}.
The energy spectra are estimated over the central Arctic region with $128\times128$ grid points and averaged in logarithmic space over all initialisation times.
To estimate the extent accuracy, the predicted sea-ice concentration is converted into an extent mask (concentration $>0.15$).
The accuracy is then the agreement between the prediction and the truth over all ocean grid points.

Using the root mean-squared error (RMSE) as metric, GenSIM's ensemble mean outperforms these baselines (Fig. \ref{fig:main_short_forecast}a).
Across all six predicted variables, GenSIM improves the scores, reaching an improvement of around $20\,\%$ compared to the deterministic forecasts.
In the central Arctic, which is mostly year-long covered by sea ice, GenSIM achieves substantial improvements, and remarkably also in regions with marginal ice zones (MIZ), though small deficiencies in the sea-ice concentration during the change from melting to refreezing remains (see also Fig. \ref{fig:app_scorecard_a} and Fig. \ref{fig:app_scorecard_b}).
These results are striking: while the deterministic model is specifically trained to minimise the RMSE, GenSIM--designed as generative flow--achieves an indirect reduction.
Derived from independent predictions, GenSIM's ensemble mean is a true statistical average, explaining why it can outperform the deterministic model \cite{bonavita_limitations_2024}.
Although well-tuned for 12 hours, GenSIM's ensemble becomes seemingly underdispersive with increasing lead time, leading to typical spread-skill ratios of $0.8$ after 10 days.
This underdispersion could be remedied by injecting more stochasticity \cite{price_probabilistic_2025} or tuning on the validation dataset \cite{couairon_archesweather_2024} from which we refrain.

Like the geophysical model neXtSIM, GenSIM retains the energy at all scales even after a 10-day forecast (Fig. \ref{fig:main_short_forecast}c), while the ensemble mean forecast and the deterministic model lose small-scale energy.
This loss of small-scale information effectively smooths the forecast out \cite{durand_datadriven_2024, finn_generative_2024}, leading to physical inconsistencies \cite{bonavita_limitations_2024} and even instabilities in long-term simulations.
GenSIM additionally predicts the sea-ice extent more accurately, which becomes especially evident during summer and autumn (Fig. \ref{fig:main_short_forecast}d).
GenSIM reprents better the sea-ice edge and marginal ice zone (MIZ), which we partially attribute to the generative nature of the model and partially to the novel censored loss function that explicitly incorporates the physical bounding of the variables.
In conclusion, GenSIM outperforms the deterministic model and resolves the problems with the loss of small-scale information, like similarly found for the atmosphere \cite{price_probabilistic_2025}, while better constraining the MIZ.

\subsubsection*{Score cards for short-term predictions}\label{app:scorecards}

\begin{figure}[ht]
    \centering
    \includegraphics[width=\textwidth]{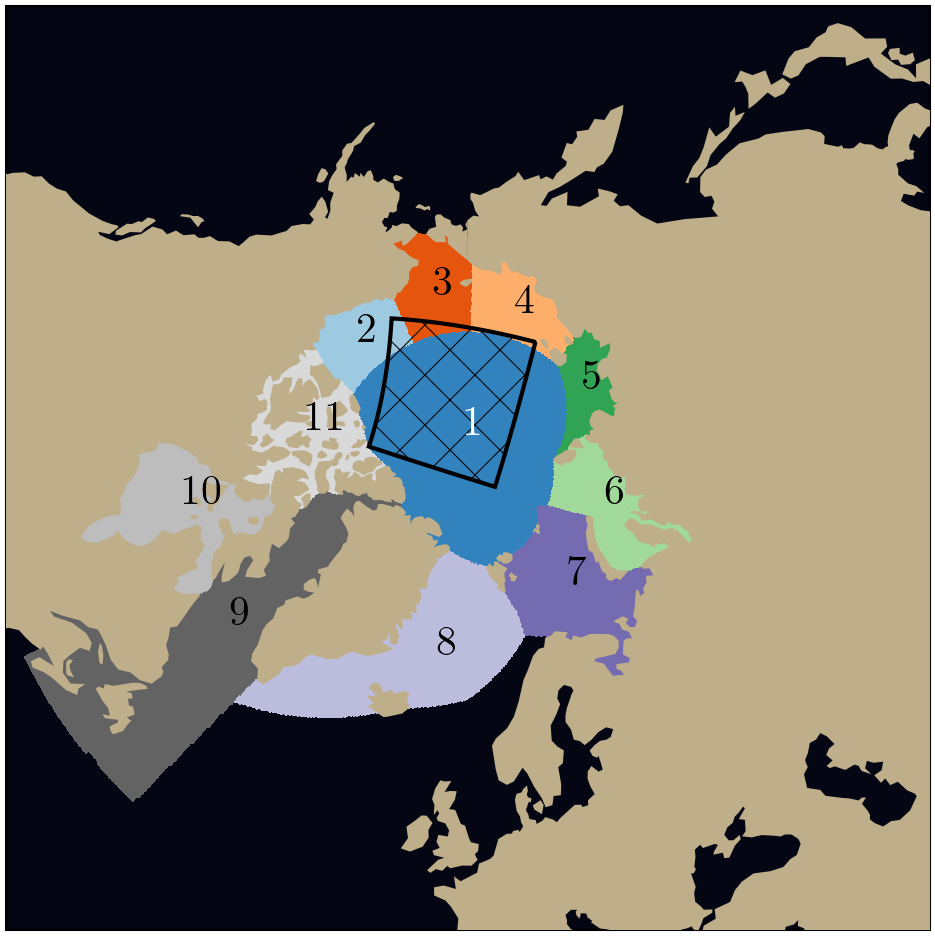}
    \caption{
        The different regions used throughout the manuscript with the following names: 1=central Arctic; 2=Beaufort sea; 3=Chukchi sea; 4=east Siberian sea; 5=Laptev sea; 6=Kara sea; 7=Barents sea; 8=east Greenland sea; 9=Baffin bay, Labrador seas, and gulf of St. Lawrence; 10=Hudson bay; 11=Canadian archipelago.
        The black hatched rectangle is the region in the central Arctic used to estimate the spectra and fractal scaling.
        The regions are extracted from NSIDC's regions on the EASE-Grid 2.0 with a 3.125 km resolution \cite{meier_arctic_2023}.
    }\label{fig:app_regions}
\end{figure}

\begin{figure}[ht]
    \centering
    \includegraphics[width=1.\textwidth]{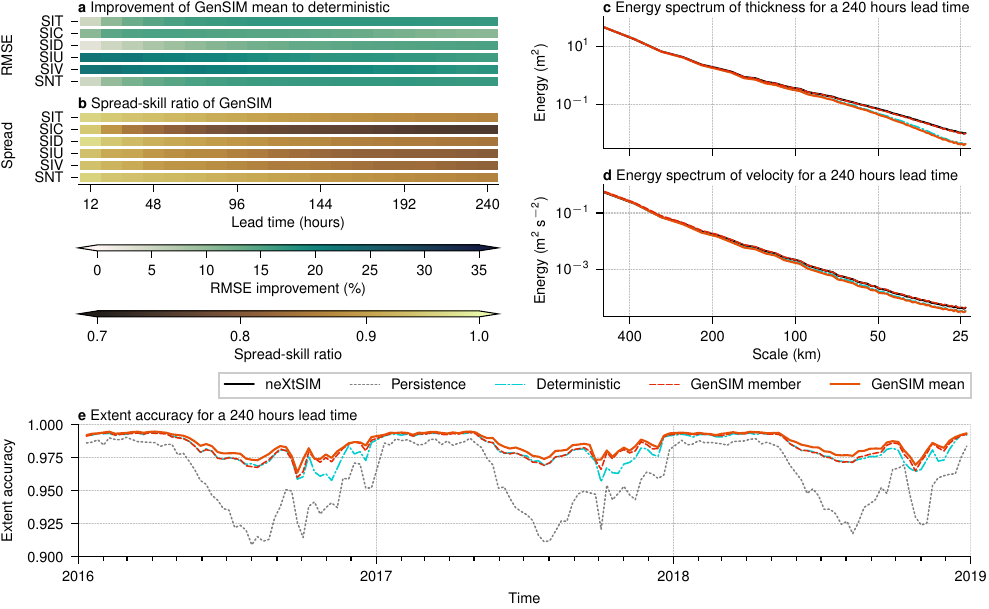}
    \caption{
        \textbf{Overview of GenSIM's short-term forecast qualities.}
        Shown are: (a) the improvement in the root-mean-squared error (RMSE) of GenSIM's ensemble mean compared to the deterministic model and the ratio between the ensemble standard deviation to the RMSE of GenSIM's ensemble mean, both are averaged over all ice-covered grid points (sit=thickness, sic=concentration, sid=damage, siu=$u$-drift, siv=$v$-drift, snt=snow-on-ice thickness);
        (b) the 10-day RMSE improvement of GenSIM's ensemble mean to the deterministic model split up into regions, averaged across all six predictive variables after normalizing with their climatology; (c) the 10-day spectrum of the sea-ice thickness in the central Arctic, averaged in log-space across all forecasts; (d) accuracy of the 10-day forecast for the sea-ice extent with the median accuracy for the ensemble members.
        The used regions for (b) and (c) are depicted in Fig. \ref{fig:app_regions}.
    }\label{fig:main_short_forecast}
\end{figure}

Inspired by ECMWF's score cards (\url{https://sites.ecmwf.int/ifs/scorecards/}, last access: 19th August 2025), the root-mean-squared error is decomposed into averages across seasons and regions.
For that decomposition, we use the indexed regions as indicated in Fig. \ref{fig:app_regions}.
Mesh cells over the Atlantic ocean that mostly cover open water are omitted from the estimation of the root-mean-squared error (RMSE).
As previously, the RMSE is estimated only over ice-covered mesh cells (concentration $>0.15$).
The score cards (Fig. \ref{fig:app_scorecard_a} and Fig. \ref{fig:app_scorecard_b}) show how GenSIM's ensemble mean improves the predictions compared to the deterministic surrogate model.
Note that there is almost no snow during summer (JAS), making its error estimate unreliable in that season.

Across all seasons and all regions, GenSIM's ensemble mean shows a persistent improvement in the RMSE compared to the deterministic surrogate model.
However, when sea ice starts to refreeze during autumn (OND), GenSIM's sea-ice concentration forecast has apparently some issues, which lead to a degradation of the forecasts over lead times of several days, generally impacting the scores for the sea-ice concentration.
Potentially, a tuning of the sampling scheduling for GenSIM could remedy this degradation.

\subsubsection*{Scaling analysis}\label{app:scaling_analysis}

In addition to the resolved dynamics, sea ice is subject to sub-grid dynamics that act at scales below the resolved $12\,\text{km}$ resolution with a brittle behaviour.
For a qualitative analysis, we imitate how features are tracked and deformations rates are calculated with the Environment and Climate Change Canada automated sea ice tracking system (ECCC-ASITS) based on Sentinel-1 synthetic aperture radar (SAR) imagery \cite{plante_sea_2025} (data available at \url{https://zenodo.org/records/13936609}, last access: 5th November 2025).
We use the individual feature pairs and their triangulation found by ECCC-ASITS from 2018-02-16 03:00 to 2018-02-25 15:00.
Based on their temporal and spatial difference, we estimate the features' drift by the relation
\begin{equation}
    \bar{\mathbf{u}}_{t} = \frac{\mathbf{x}_{t}-\mathbf{x}_{s}}{t-s}\label{eq:feature_drift},
\end{equation}
where $\mathbf{x}_{t}$ is the position at time $t$ and $\mathbf{x}_{s}$ the position at time $s$ with $t > s$.
We use the original per satellite triangulation of the features and retain all triangles where the original signal-to-noise ratio is above $0.5$. 
The drift gradients, needed for the deformation rates, are estimated based on the triangles with the contour integral \cite{rampal_multifractal_2019}, e.g., $u_{x} = \frac{1}{A}\oint u\,dy$ for the $u$-drift gradient in $x$-direction with $A$ as triangle area.
The shear deformation, $\dot{\epsilon}_{\mathrm{shear}}$, divergent deformation, $\dot{\epsilon}_{\mathrm{div}}$, and total deformation rates $\dot{\epsilon}_{\mathrm{tot}}$ are estimated by
\begin{align}
    \dot{\epsilon}_{\mathrm{shear}} &= \sqrt{(u_{x}-v_{y})^2 + (u_{y}+v_{x})^2},\label{eq:deform_shear}\\
    \dot{\epsilon}_{\mathrm{div}} &= u_{x}+v_{y},\label{eq:deform_div}\\
    \text{and } \dot{\epsilon}_{\mathrm{tot}} &= \sqrt{\dot{\epsilon}^2_{\mathrm{shear}} + \dot{\epsilon}^2_{\mathrm{div}}}.\label{eq:deform_total}
\end{align}

Every 12 hours, we initialise particles at positions of the found features in a $\pm$ 6-hour window.
Based on the simulated sea-ice drift in the GenSIM and neXtSIM simulations, we advect all particles forward to approximately match the same period ($\pm$ 6 hours) as used for the original features.
As advection scheme, we use a simple forward Euler integration with an hourly time step, a bilinear drift interpolation in space, and linear interpolation in time.
The averaged drift and deformation rates for GenSIM and neXtSIM are then estimated based on these advected particles.

\begin{figure}[ht!]
    \centering
    \includegraphics[width=1.\textwidth]{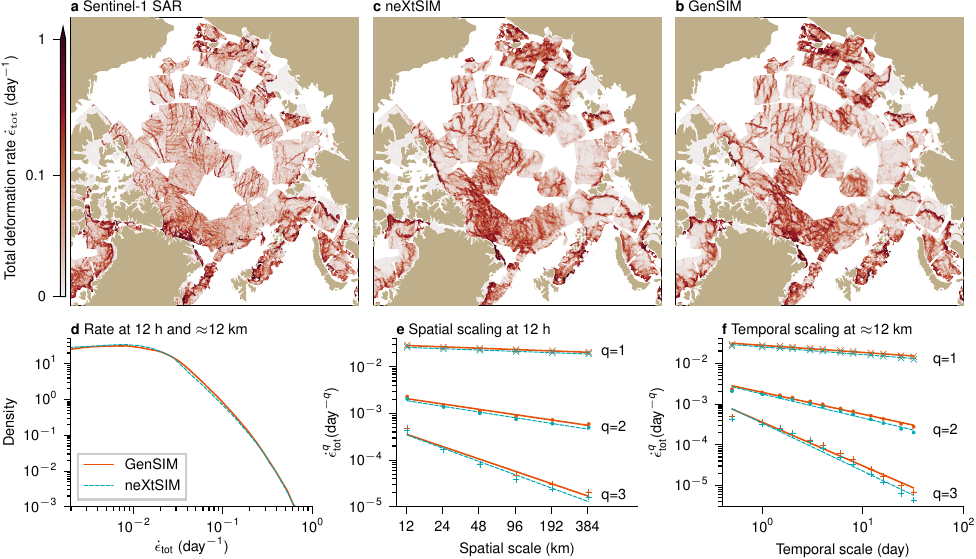}
    \caption{
        \textbf{Fractal scaling of GenSIM's dynamics.} Triangulated total deformation rates as inferred from Sentinel-1 synthetic aperture radar (SAR) imagery (a), from advecting with neXtSIM's predicted drift (b), and from advecting with GenSIM's predicted drift (c) for the period 2018-02-16 03:00--2018-02-25 15:00.
        The deformation rates are estimated at the original data resolution and image pairs with an averaged signal-to-noise ratio $<0.5$ are filtered out.
        Note that the rates are temporally ordered with the newest images on top.
        (d) Distribution of total deformation rates for GenSIM and neXtSIM at the nominal resolution of around 12 km and an averaging window of 12 hours.
        (e) First three distributional moments of the total deformation rate (q) with coarse-grained spatial scale and the fitted power laws as linear lines for an averaging window of 12 hours.
        (f) First three distributional moments of the total deformation rate (q) with increasing averaging window at the nominal resolution of around 12 km.
        In (d)--(f), the total deformation rates are estimated based on continuously tracked particles from 2017-12-01 03:00 to 2018-03-31 15:00, which were advected with the drift of GenSIM and neXtSIM, respectively.
        The particles were initialised in a $65\times65$ mesh at the cell centres of an area in the central Arctic.
        The power laws in (e) and (f) are fitted based on the relations $\left\langle\dot{\epsilon}^{q}_{\mathrm{tot}}(L)\right\rangle \sim L^{-\beta(q)}$ and $\left\langle\dot{\epsilon}^{q}_{\mathrm{tot}}(T)\right\rangle \sim T^{-\alpha(q)}$, respectively \cite{rampal_multifractal_2019}.
    }\label{fig:deformation}
\end{figure}

GenSIM demonstrates its strong skill to reproduce key brittle characteristics of observed Arctic sea-ice deformation from Sentinel-1 SAR images, while performing comparably to neXtSIM (Fig.~\ref{fig:deformation}a--c).
GenSIM successfully captures the spatial organisation and intensity of the deformation features from SAR and neXtSIM, including anisotropic features and high-strain bands, even though neXtSIM and GenSIM have more diffuse structures due to their spatial and temporal averaging (the predicted drift is six-hour averages). 

In neXtSIM, the sub-grid dynamics are parametrised by the brittle Bingham-Maxwell rheology \cite{olason_new_2022}, thanks to which the model can reproduce multifractal resolved short-term dynamics as observed by satellites.
To quantitatively compare the scaling of GenSIM's and neXtSIM's short-term dynamics in space and time, we initialise an ensemble of $65\times65$ particles at the cell centres of the curvilinear mesh in the central Arctic, far from the coast, at 2017-12-01 03:00.
As the curvilinear mesh is structured, the particles naturally form quadrangles.
These particles are then continuously advected with the predicted drift from GenSIM and neXtSIM based on the same advection scheme as described before.
We track the positions of these advected particles from 2017-12-01 03:00 to 2018-03-31 15:00.
The averaged drift is then calculated based on the difference between two corresponding positions in time with Eq.~\ref{eq:feature_drift}.

To estimate the temporal and spatial scaling, we coarse-grain by calculating averaged drifts at various time periods and by averaging drift gradients, respectively.
The drift gradients from the original quadrangles are merged into larger quadrangles with an area-weighted averaging \cite{bouchat_reassessing_2020}.
With these larger quadrangles and their averaged gradients, we estimate the deformation rates at nominal temporal scales ${T \in [12\,\text{h}, 1\,\text{d}, 2\,\text{d}, 4\,\text{d}, 8\,\text{d}, 16\,\text{d}, 32\,\text{d}]}$ and at nominal spatial scales ${L \in [12.25\,\text{km}, 24.5\,\text{km}, 49\,\text{km}, 98\,\text{km}, 196\,\text{km}, 392\,\text{km}]}$.
For each spatial scales, we require that $2^{-1/2} L \le l \le 2^{1/2} L$ holds for the area of the merged quadrangles $l$.

At each scale, we track the first three distributional moments of the deformation rates.
The scaling analysis then assumes power laws in space and time such that the moments $q$ of the averaged rates at scale $L$ and $T$ follow $\big\langle \dot{\epsilon}^{q}(L, T)\big\rangle \sim L^{-\beta(q)}$ and $\big\langle \dot{\epsilon}^{q}(L, T)\big\rangle \sim T^{-\alpha(q)}$, respectively.
The scaling coefficients $\beta(q)$ and $\alpha(q)$ in space and time, respectively, describe how the moments change by coarse-graining.
The dependence of the scaling coefficients on the moment gives an estimate how much the localisation of fracturing events are dependent on the magnitude of these events, a positive curvature indicates thereby a multifractal and intermittent behaviour \cite{rampal_multifractal_2019, bouchat_reassessing_2020}.

Here, GenSIM maintains the behaviour of the sea-ice dynamics as modelled by neXtSIM: the heavy-tailed distribution of total deformation rates from GenSIM mirrors neXtSIM's at the 12 km and 12 hour scale (Fig.~\ref{fig:deformation}d).
The spatial scaling of GenSIM's dynamics closely aligns with neXtSIM's across orders of magnitude (Fig.~\ref{fig:deformation}e), even though GenSIM has a small positive bias due to an overestimation of intermediate deformation rates.
Moreover, GenSIM yields a similar intermittency to neXtSIM across a range of temporal averaging windows (Fig.~\ref{fig:deformation}f).
Despite generally matching neXtSIM, GenSIM slightly underestimates the scaling coefficients at higher moments, as indicated by the slopes of the fitted power laws for $q=3$.
Consequently, GenSIM shows a multifractal spatial and temporal scaling to a slightly lesser extent than neXtSIM.
Nevertheless, the scaling invariance of GenSIM is remarkable, considering that it is estimated based on diagnosed deformation rates, which require the right correlations in space and time between the predicted drift components.
This underlines once again GenSIM's physical consistency, which allows us to perform such advanced analyses.

\subsubsection*{Sea-ice extent}\label{app:iceindex}

The sea-ice extent is evaluated similarly to the sea-ice volume in the main manuscript.
All simulated and retrieved sea-ice concentration products are bilinearly interpolated to the EASE-25-km mesh from OSISAF and averaged to a monthly timescale.
The monthly averaged sea-ice concentration is thresholded by the usual threshold of $\ge0.15$.

\begin{figure}[ht]
    \centering
    \includegraphics[width=0.8\textwidth]{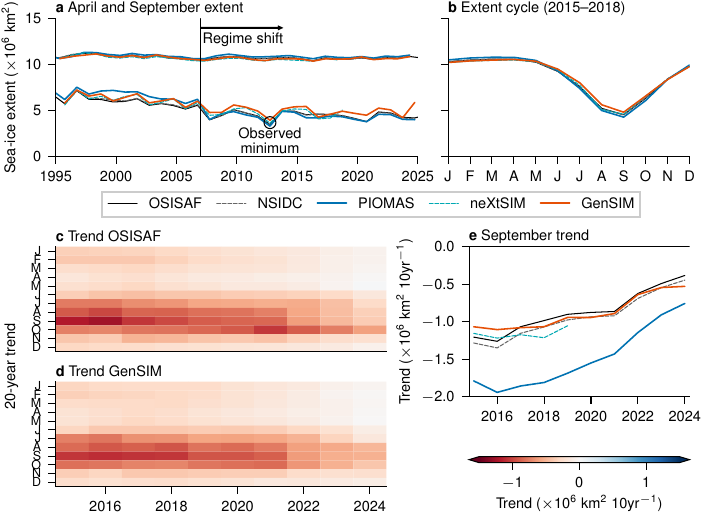}
    \caption{
        \textbf{GenSIM predicts the slowing in the decrease of the sea-ice extent.}
        (a, b) Monthly averaged sea-ice extent for April and September (a) and seasonal cycle of the sea-ice extent averaged for 2015--2018 (b), both inferred for satellite products from OSISAF and NSIDC, the PIOMAS reanalysis, and the simulations from neXtSIM and GenSIM.
        (c--e) Sea-ice extent trend over 20 years in the OSISAF product (c) and in the GenSIM simulation (d) for different months, as well as the September trend for all products.
    }\label{fig:app_extent}
\end{figure}

For April and September, all products have a similar evolution of the sea-ice extent (Fig.~\ref{fig:app_extent}a) with small positive biases in September for PIOMAS before 2007 and for GenSIM after 2007.
This positive bias in GenSIM is also evident in the seasonal cycle of the extent for the period 2015--2018 (Fig.~\ref{fig:app_extent}b).
Nonetheless, GenSIM's 20-year trend for individual months is remarkably similar to OSISAF's.
GenSIM successfully shows the slowing in the decrease of the sea-ice extent \cite{england_minimal_2025} (Fig.~\ref{fig:app_extent}c--e).
This becomes especially striking when compared to the other products for the 20-year trend in September.
Therefore, GenSIM simulates the Arctic-wide sea-ice extent as observed, suggesting its reliability for longer climate-like projections.

\subsubsection*{Sea-ice thickness and area}\label{app:thickness_area}

As the sea-ice volume is the product of the sea-ice thickness with the sea-ice concentration, we evaluate these two components independently with the averaged sea-ice thickness and the sea-ice area.
Like previously, all fields are bilinearly interpolated to the common EASE-25-km mesh and averaged to monthly values.
The thickness of a grid cell is only used for the averaged thickness, if its averaged concentration is above $0.15$.

\begin{figure}[ht]
    \centering
    \includegraphics[width=0.8\textwidth]{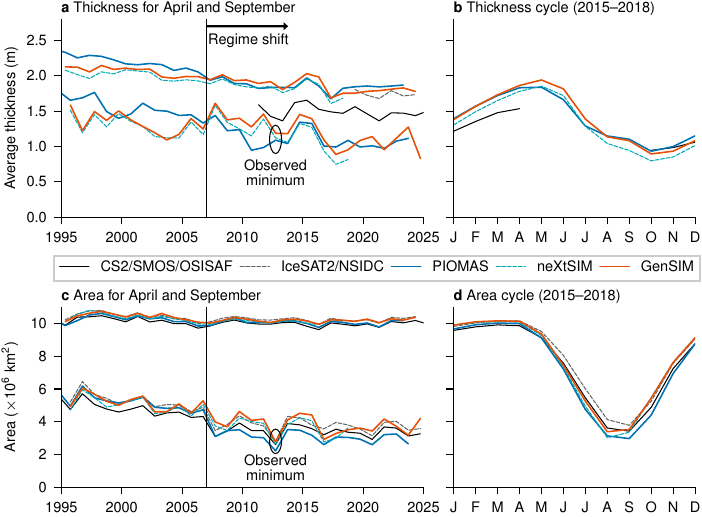}
    \caption{
        \textbf{GenSIM captures the evolution of sea-ice thickness and area.}
        (a--b) Monthly averaged sea-ice thickness for April and September (a) and seasonal cycle averaged for 2015--2018 (b).
        (c--d) Monthly averaged sea-ice area for April and September (c) and seasonal cycle averaged for 2015--2018 (d).
        Please note that the CS2/SMOS and IceSAT2 product for the sea-ice thickness are only available for the winter months, while the OSISAF and NSIDC products for sea-ice area are available over the full years.
    }\label{fig:app_thickness_area}
\end{figure}

GenSIM follows the trend in the sea-ice thickness towards thinner ice remarkably well and matches neXtSIM with a small positive bias (Fig.~\ref{fig:app_thickness_area}a).
Before 2007, both GenSIM and neXtSIM have the tendency towards thinner ice than inferred from PIOMAS.
Yet, after 2007, GenSIM's average thickness captures the annual variations in PIOMAS and the IceSAT2 satellite product.
This agreement becomes especially evident in April.
GenSIM and neXtSIM have both a sea-ice thickness minimum in September 2017 and 2018, which also drives their minimum for the sea-ice volume.
While the small bias of GenSIM compared to neXtSIM manifests in the annual cycle, it fits the cycle of PIOMAS from August to April better (Fig.~\ref{fig:app_thickness_area}b).
Compared to all other products, even to IceSAT2, CS2/SMOS exhibits a neagtive bias in the thickness that increases with increasing freezing period.
Similar to the sea-ice extent, the averaged area of GenSIM is in the range of all other products, include the OSISAF and NSIDC climate data records (Fig.~\ref{fig:app_thickness_area}c\&d).
Hence, these results suggest that existing differences and biases are mostly due to differences in the sea-ice thickness while the sea-ice extent and area are rather well constrained.

\subsubsection*{Diagnosed sea-ice age}\label{app:iceage}

The sea-ice age is diagnosed as incremental tracer variable, which is advected with the predicted drift and retained over ice-covered mesh cells.
To advect in Eulerian space, we use a backward semi-Lagrangian integration scheme \cite{germann_scaledependence_2002, finn_generative_2024}:
all unmasked cell centres are advected $12\,\text{h}$ backward in time with an hourly time step.
The predicted drift is mapped to advected centres with a nearest neighbour interpolation in space and a linear interpolation in time.
After advecting the centres, each unmasked cell with its centre, $(x_{t}, y_{t})$, has a corresponding displaced centre $12$ hours earlier, $(x'_{t-12\,\text{h}}, y'_{t-12\,\text{h}})$.
To obtain the advected sea-ice age, we bilinearly interpolate the gridded sea-ice age, $\mathrm{age}_{t-12\,\text{h}}(x'_{t-12\,\text{h}}, y'_{t-12\,\text{h}})$.
The new sea-ice age reads then
\begin{align}
    \mathrm{age}_{t}(x_{t}, y_{t}) =& \begin{cases}
        \mathrm{age}_{t-12\,\text{h}}(x'_{t-12\,\text{h}}, y'_{t-12\,\text{h}}) + 12\,\text{h}, &\text{if } \mathrm{conc}_{t}(x_{t}, y_{t}) > 0.15\\
        0, &\text{otherwise}.
    \end{cases}
\end{align}
Effectively, the sea-ice age is reset if the mesh cell is no longer covered by ice (concentration $<0.15$) or if the advected age comes from land.
Caused by the bilinear interpolation within the advection scheme, the predicted sea-ice age is exposed to some diffusion.

To assess the impact of the warming climate on the predicted sea-ice age, we advect the sea-ice age from 2000-01-01 03:00 to 2003-12-31 15:00 and from 2015-01-01 03:00 to 2018-12-31 15:00 for GenSIM and neXtSIM.
Hence, the upper bound for the sea-ice age at the end of these windows is four years.
These estimates are compared to the satellite sea-ice age product from the National Snow and Ice Data Center (NSIDC) \cite{tschudi_easegrid_2019} on the EASE-12.5-km mesh.

\begin{figure}[ht]
    \centering
    \includegraphics{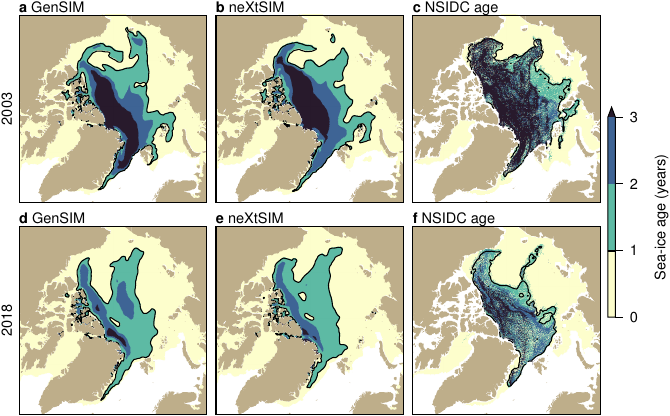}
    \caption{
        \textbf{GenSIM simulates the decrease of the sea-ice age.}
        (a--c) Sea-ice age for GenSIM (a) for neXtSIM (b), and as observed (c) at 2003-12-31 15:00 UTC.
        (d--f) Sea-ice age for GenSIM (d), for neXtSIM (e), and as observed (f) at 2018-12-31 15:00 UTC.
        The black contour lines indicate multi-year ice.
        Note that the contour lines for the observed age in (c) and (f) are smoothed with a $4\times4$ averaging filter for visualisation purposes.
        The sea-ice age is post-processed from the simulated sea-ice drift and sea-ice concentration, while the observations are the \textit{EASE-Grid Sea Ice Age} product from the NSIDC \cite{tschudi_easegrid_2019}.
    }\label{fig:app_ice_age}
\end{figure}

Remarkably, GenSIM, neXtSIM and the satellite product agree quite well in the large-scale distribution of the sea-ice age for 2003 (Fig.~\ref{fig:app_ice_age}a--c), as well as 2018 (Fig.~\ref{fig:app_ice_age}d--f).
Also, all products show a transition from large fractions of multi-year ice in 2003 to larger fractions in single-year ice or younger in 2018.
To mirror the observed sea-ice age, the models needs to represent the correct evolution and correlations in the sea-ice drift and sea-ice concentration as similarly observed.
Hence, GenSIM can represent them like neXtSIM does, suggesting that GenSIM performs simulations that have similar quality as those from neXtSIM.

\subsubsection*{Generalisation to other resolutions}\label{app:resolution}

GenSIM is designed to work in resolutions between 12 km and 100 km by randomly averaging the data during training.
To test the capabilities of GenSIM at other resolutions than the original 12 km, we perform an experiment for ten years at 96 km.
Initialised with a $8\times8$ coarse-grained predictions from the high-resolution GenSIM run on 2015-01-01 03:00, the model is run until 2024-12-31 15:00 with coarse-grained ERA5 forcings.
This low-resolution simulation is compared to the high-resolution simulation and the PIOMAS reanalysis.

For all three products, the sea-ice volume is estimated by bilinearly interpolating onto a common EASE-100-km mesh.
As for the original volume estimation, a common mask is used and all data is monthly averaged.
Please note that the sea-ice volume estimates are below the ones from Fig.~\ref{fig:main_long_forecast}, as a coarser masking has been used.

\begin{figure}[ht]
    \centering
    \includegraphics{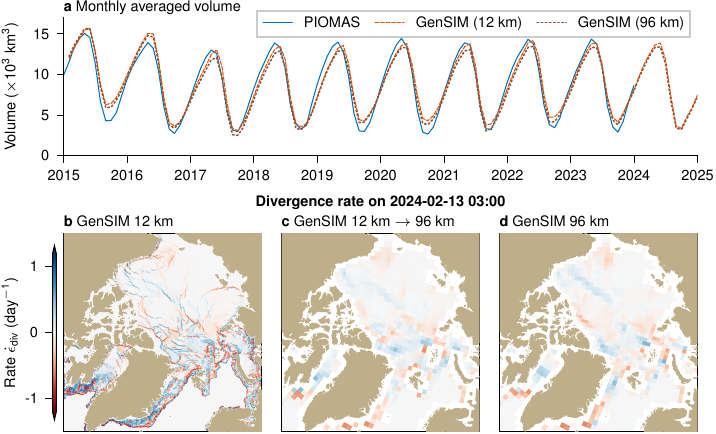}
    \caption{
        \textbf{GenSIM performs similarly across resolutions.}
        (a) Monthly averaged sea-ice volume on an EASE-100 km mesh for the PIOMAS reanalysis and simulations from GenSIM at the original resolution ($12\,\text{km}$) and at coarse resolution ($96\,\text{km}$).
        Note that the volume estimates are smaller than in Fig.~\ref{fig:main_long_forecast} due to a coarse-grained masking.
        (b--d) Divergence rate for 2024-02-13 03:00 estimated with the drift from GenSIM at the original resolution (b), the drift coarse-grained from $12\,\text{km}$ to $96\,\text{km}$ (c), the drift from GenSIM at coarse resolution (d).
    }\label{fig:app_coarse}
\end{figure}

The low-resolution simulation follows remarkably closely the previously established performance of GenSIM at high resolutions (Fig.~\ref{fig:app_coarse}a).
The bias of GenSIM at high resolutions compared to PIOMAS is slightly reduced for the low-resolution simulation, while the seasonal cycle is matched almost perfectly.
The representation of thermodynamics in GenSIM apparently works even at coarse resolutions, and we would expect similar results at other resolutions within the trained scales

We analyse the dynamics with the same prediction time as for Fig~\ref{fig:consistency} by estimating the divergent deformation rate on 2024-02-13 03:00.
In addition to estimating the deformation rate at the high resolution (Fig.~\ref{fig:app_coarse}b) and low resolution (Fig.~\ref{fig:app_coarse}d), we also coarse-grain the drift from the high resolution to the low resolution, with which we then estimate the deformation rates (Fig.~\ref{fig:app_coarse}c).

By coarse-graining the drift from the high resolution, we attenuate the deformation rates and average out small-scale variations such that only large-scale patterns remain.
However, when run at low resolutions, GenSIM can suddenly resolve features from the high-resolution simulation which were strongly averaged out during the coarse graining, e.g., the divergent lead in the crossing the central Arctic from the south-east to the north-west.
Consequently, at the low resolution, GenSIM exhibits dynamics as we would similar expect them to behave if the sub-grid scales are properly parametrised.

These results suggest that GenSIM learns to generalise its dynamics and thermodynamics at different resolutions.
Hence, the strategy of training the model with randomly coarse-grained data was successful and is apparently a way to artificially increase the data amount.
Furthermore, this renders GenSIM as a highly flexible sea-ice model, indicating its potential to replace sea-ice component models in coupled Earth system models.

\subsubsection*{Mathematical framework for surrogate modelling}\label{sec:app_mathematics}

Our goal is to develop a powerful surrogate model that forecasts sea-ice states with a 12-hour lead time, $\mathbf{x}_{t+12\,\text{h}}$, given some initial states $\mathbf{x}_{t}$ and external forcings $\mathbf{F}_{t:t+12\,\text{h}}$ provided at initial time $t$ and 12 hours later, $t+12\,\text{h}$.
As a reminder, the predictive variables are the sea-ice thickness, the concentration, the damage, the two drift components, and the snow-on-ice thickness.
Our model predicts the tendency within the 12 hours,
\begin{equation}
    \mathbf{x}_{t+12\,\text{h}} \approx \widehat{\mathbf{x}}_{\boldsymbol{\theta}, t+12\,\text{h}} = \mathbf{x}_{t} + \mathcal{M}_{\boldsymbol{\theta}}(\mathbf{x}_{t}, \mathbf{F}_{t:t+12\,\text{h}}),\label{eq:prediction_nn}
\end{equation}
with $\widehat{\mathbf{x}}_{\boldsymbol{\theta}, t+12\,\text{h}}$ as the model prediction.
The tendency model $\mathcal{M}_{\boldsymbol{\theta}}(\cdot)$ is based on a deep neural network with its learnable weights and biases $\boldsymbol{\theta}$.
Here, we train two different types of models:
a deterministic model \cite{durand_datadriven_2024} and a generative model based on flow matching \cite{liu_flow_2022, albergo_stochastic_2023, lipman_flow_2023}.

To train the models, we sample data triplets from a training dataset $(\mathbf{x}_{t}, \mathbf{x}_{t+12\,\text{h}}, \mathbf{F}_{t:t+12\,\text{h}}) \sim \mathcal{D}$ and minimise the loss functions using variants of stochastic gradient descent.
Once trained, we employ the models as outlined in~Equation~\ref{eq:prediction_nn} to propagate the initial states forward and to generate a 12-hour lead time prediction.
For predictions extending beyond 12 hours, we utilise the model autoregressively, wherein the preceding prediction serves as the input for the subsequent time step, with the forcings being updated accordingly.

In practice, states are processed in a normalised space, in which the output of the neural network is scaled.
The latent state that the neural network outputs reads
\begin{equation}
    \mathbf{z} = \frac{\mathbf{x}_{t+12\,\text{h}}-\mathbf{x}_{t}}{\boldsymbol{\sigma}_{\textrm{clim}}}.\label{eq:latent_space}
\end{equation}
The normalisation factor $1/{\boldsymbol{\sigma}_{\textrm{clim}}}$ is the reciprocal of the per-variable climatological standard deviation, estimated over all training samples and all unmasked ocean grid points.
To recover the prediction, we can scale and add the output to the initial states,
$\widehat{\mathbf{x}}_{t+12\,\text{h}} = \mathbf{x}_{t} + \boldsymbol{\sigma}_{\textrm{clim}}\widehat{\mathbf{z}}$.

The deterministic model is trained to predict the mean of an assumed Gaussian distribution as a single outcome, conditioned on the initial states and forcings.
For the autoregressive prediction, the mean is treated as point estimate, which is reused as input for the next time step.

For training, we assume that true sea-ice state after 12 hours is drawn from a Gaussian distribution with the prediction as the mean and a constant diagonal covariance matrix,
\begin{equation}
    \mathbf{x}_{t+12\,\text{h}} \sim \mathcal{N}\left(\widehat{\mathbf{x}}_{\boldsymbol{\theta}, t+12\,\text{h}}, \boldsymbol{\sigma}_{\textrm{clim}}^2 \mathbf{I}\right),
\end{equation}
with the values on the diagonal of the covariance as the vector of all climatological standard deviations.
Using maximum likelihood estimation for the predicted mean, the corresponding loss function is a weighted mean-squared error,
\begin{equation}
    \mathcal{L}_{\textrm{det}}(\boldsymbol{\theta}) = \mathbb{E}_{\mathcal{D}}\left[\mathbf{w}\lVert \mathbf{x}_{t+12\,\text{h}}-\widehat{\mathbf{x}}_{\boldsymbol{\theta}, t+12\,\text{h}}\rVert^{2}_{2}\right],\label{eq:mse_clim_loss}
\end{equation}
with $\mathbf{w} = 1/{\boldsymbol{\sigma}_{\textrm{clim}}^2}$ as weighting vector.
This loss function is commonly used for other surrogate models \cite{lam_graphcast_2023, lang_aifs_2024, durand_datadriven_2024} and serves as a baseline for our generative model.

In normalised space, the loss function reduces equivalently to a mean-squared error between the targetted latent variable and the predicted latent variable, as the weighting is subsumed into the normalisation,
\begin{equation}
    \mathcal{L}_{\textrm{det}}(\boldsymbol{\theta}) = \mathbb{E}_{\mathcal{D}} \left[\lVert \mathbf{z}-\widehat{\mathbf{z}}_{\boldsymbol{\theta}}\rVert^{2}_{2}\right].
\end{equation}
This underlines the role of the climatological standard deviation in the deterministic model: it is used as scaling for the output of the neural network but also as weighting in the loss function.

While the deterministic model predicts a single outcome, our goal with a model based on generative flows is to produce a valid sample from the predictive probability distribution through refinement.
Starting from common flow matching, we derive this refinement process, and we further introduce the concepts needed for the loss functions to train the generative model.
In the end, we will derive censored flow matching as a way to constrain the resulting generative flow by physical bounds.

In our flow matching model, the latent state $\mathbf{z}_{\tau}$ is evolved in pseudo time $\tau$ from pure noise at $\tau=0$ to a sample from the targetted distribution at $\tau=1$.
Note that the latent state is dependent on the pseudo time $\tau$ as denoted by its subscript, while the true sea-ice state depends on real time $t$.
The prior distribution of the latent state is a Gaussian with zero mean and identity covariance,
\begin{equation}
    \mathbf{z}_{0} \sim \mathcal{N}(\boldsymbol{0}, \mathbf{I}).\label{eq:flow_prior}
\end{equation}
During training, the targetted distribution is known through samples, $\mathbf{z}_{1}=(\mathbf{x}_{t+12\,\text{h}}-\mathbf{x})/\boldsymbol{\sigma}_{\textrm{clim}}$, as previously.

The ordinary differential equation (ODE) that governs the generative flow in normalised space is given by
\begin{equation}
    \frac{\mathrm{d}\mathbf{z}_{\tau}}{\mathrm{d}\tau} = \mathbf{u}(\mathbf{z}_{\tau}),\label{eq:ode_drift}
\end{equation}
where $\mathbf{u}(\mathbf{z}_{\tau})$ represents the velocity of the flow.
When the velocity is known, we can integrate the ODE in pseudo-time and produce new data samples.

For training, the flow is constructed by a linear interpolant,
\begin{equation}
    \mathbf{z}_{\tau} = \tau \mathbf{z}_{1} + (1-\tau) \mathbf{z}_{0},\label{eq:linear_flow}
\end{equation}
resulting in a constant true velocity independent of $\mathbf{z}_{\tau}$ and $\tau$,
\begin{align}
    \mathbf{u} = \mathbf{z}_{1}-\mathbf{z}_{0}.\label{eq:flow_velocity}
\end{align}
Since $\mathbf{z}_{1}$ is targetted and the analytical velocity therefore unknown during prediction, we train the neural network with its parameters $\boldsymbol{\theta}$ to approximate the true velocity, $\mathbf{u} \approx \widehat{\mathbf{v}}_{\boldsymbol{\theta}}(\mathbf{z}, \tau, \mathbf{x}_{t}, \mathbf{F}_{t:t+12h})$.
By conditioning the approximation on the initial states and forcings, the neural network learns to sample from the conditional distribution \cite{batzolis_conditional_2021}.

To train the neural network, we draw mini-batches of samples for the pseudo time $\tau \in [0, 1]$ and the prior $\mathbf{z}_{0}\sim \mathcal{N}(\boldsymbol{0}, \mathbf{I})$ in addition to the initial and targetted states, and forcings.
After constructing the latent states $\mathbf{z}_{\tau}$, $\mathbf{z}_{1}$, the following loss function is minimised:
\begin{equation}
    \mathcal{L}_{\textrm{FM}}(\boldsymbol{\theta}) = \mathbb{E}_{\tau, \mathbf{z}_{0},\mathcal{D}} \left[ \lVert \mathbf{u}-\widehat{\mathbf{v}}_{\boldsymbol{\theta}}(\mathbf{z}_{\tau}, \tau, \mathbf{x}_{t}, \mathbf{F}_{t:t+12h})\rVert^2_{2}\right].
    \label{eq:loss_flow_matching}
\end{equation}

\begin{figure}[ht]
    \centering
    \includegraphics[width=\textwidth]{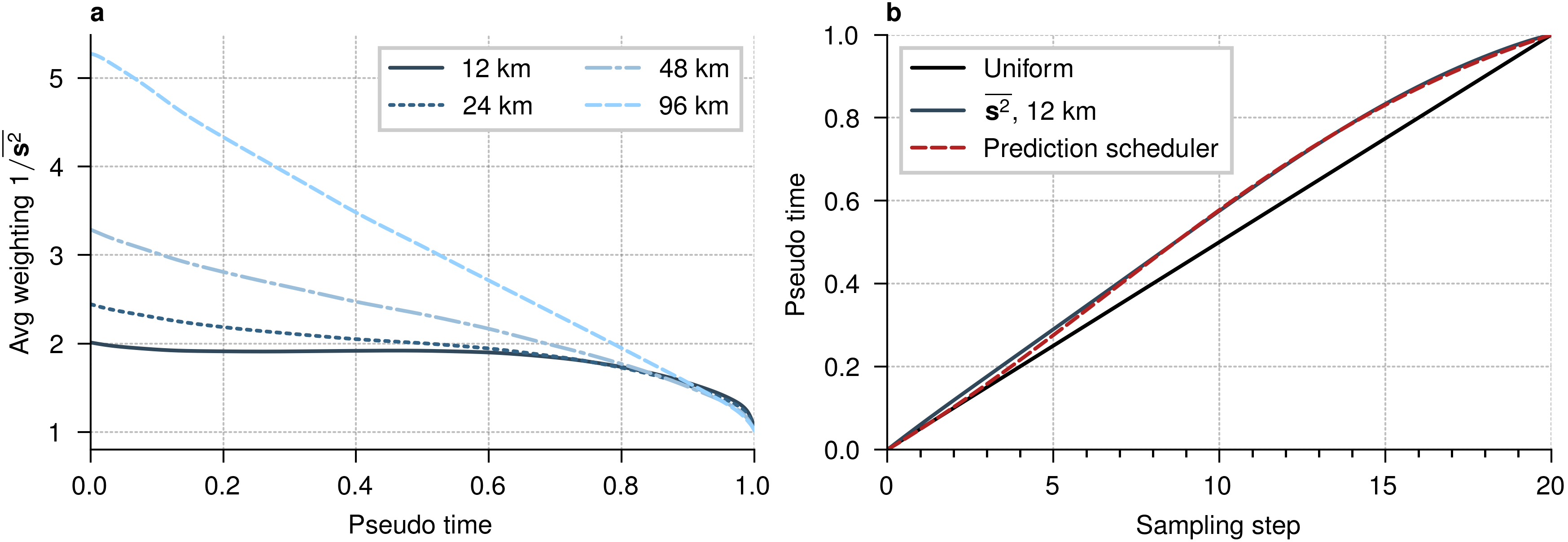}
    \caption{
        \textbf{Learned weighting of the flow matching loss}
        (a) The learned weighting, averaged across all six variables, at different resolutions indicates the importance of difference scales at different resolutions.
        (b) Used in an importance-sampling-like mechanism, the learned standard deviation can guide the tuning of the pseudo time scheduler during prediction.
        Here, the focus is on smaller scales, as opposed to a uniform scheduler.
    }\label{fig:app_weighting}
\end{figure}

Once trained, we can employ the neural network as velocity approximation for the ODE,~Equation~\ref{eq:ode_drift}.
By initialising with~Equation~\ref{eq:flow_prior} and integrating the ODE in pseudo time from $\tau=0$ to $\tau=1$, we produce a sample from the target distribution, conditioned on the initial states and forcings.
By drawing different initialisations, an ensemble of predictions can be produced from the same conditional information.
In practise, we perform the integration with a second-order Heun scheme \cite{karras_elucidating_2022} in 20 steps, where the discrete steps are scheduled with a shifted sigmoid function, as shown in Fig. \ref{fig:app_weighting}b.

Notably, if projected back into physical space, the generative flow is initialised by a sample from the prior distribution around the initial conditions $\mathcal{N}(\mathbf{x}_{t}, \boldsymbol{\sigma}^2_{\text{clim}} \mathbf{I})$ and maps to a sample from the predictive distribution for $\mathbf{x}_{t+12\,\text{h}}$.
Although a deterministic forecast, trained with~Equation~\ref{eq:mse_clim_loss}, could also be used to define the prior \cite{mardani_residual_2024,couairon_archesweather_2024}, previous work \cite{finn_generative_2024} has demonstrated negligible performance difference, and we therefore employ a persistence forecast as prior for simplicity.
Since the integration of the ODE maps from the initial conditions to the prediction, it works similarly to classical geophysical models \cite{finn_representation_2024}, integrating in pseudo time instead of real time.
Hence, we can make use of established numerical tools to render generative flow models more efficient and to scale them up, e.g., by domain decomposition.

To obtain further insights in how flow matching works and lay the ground for further concepts, we describe variational flow matching \cite{eijkelboom_variational_2024} in the following.
In the end, we demonstrate that flow matching can be seen as a generalisation of the deterministic model that refines the predictions instead of making a one-step projection.

Variational flow matching predicts the variational distribution of terminal states $q_{\boldsymbol{\theta}}(\mathbf{z}_{1} \mid \mathbf{z}_{\tau})$ instead of a point estimate for the velocity.
We can connect these two by
\begin{equation}
    \mathbf{v}_{\boldsymbol{\theta}}(\mathbf{z}_{\tau}, \tau) = \mathbb{E}_{\mathbf{z}_{1} \sim q_{\boldsymbol{\theta}}(\mathbf{z}_{1} \mid \mathbf{z}_{\tau})}\left[\mathbf{u}(\mathbf{z}_{\tau} \mid \mathbf{z}_{1})\right].\label{eq:vfm_exp_cond_velocity}
\end{equation}
The conditional velocity $\mathbf{u}(\mathbf{z}_{\tau} \mid \mathbf{z}_{1})$ is obtained from the linear interpolation,~Equation~\ref{eq:linear_flow},
\begin{equation}
    \mathbf{u}(\mathbf{z}_{\tau} \mid \mathbf{z}_{1}) = \frac{\mathbf{z}_{1}-\mathbf{z}_{\tau}}{1-\tau}.\label{eq:vfm_cond_velocity}
\end{equation}
Since this conditional velocity is analytically defined, the approximated velocity,~Equation~\ref{eq:vfm_exp_cond_velocity}, matches the true velocity, if the variational distribution matches the true distribution $p(\mathbf{z}_{1} \mid \mathbf{z}_{\tau})$, which suggests a minimisation of the Kullback-Leibler divergence between these two distributions as loss function,
\begin{align}
   \mathcal{L}_{\textrm{VFM}}(\boldsymbol{\theta}) &= \mathbb{E}_{\tau}\left[D_{\textrm{KL}}\Big(p(\mathbf{z}_{\tau})p(\mathbf{z}_{1}\mid\mathbf{z}_{\tau}) || p(\mathbf{z}_{\tau})q_{\boldsymbol{\theta}}(\mathbf{z}_{1}\mid\mathbf{z}_{\tau})\Big)\right]\notag\\
   &= -\mathbb{E}_{t, \mathbf{z}_{0}, \mathcal{D}}\left[ \log\big(q_{\boldsymbol{\theta}}(\mathbf{z}_{1}\mid\mathbf{z}_{\tau}) \big) \right] + C,
\end{align}
where $C$ represents terms that do not depend on $\boldsymbol{\theta}$.
Thanks to a mean-field approximation \cite{eijkelboom_variational_2024}, the log-likelihood can further be factorised over the $n_{\mathrm{vars}}$ variables and the $n_{\mathrm{grid}}$ grid points to
\begin{equation}
    \mathcal{L}_{\textrm{VFM}}(\boldsymbol{\theta}) = -\mathbb{E}_{t, \mathbf{z}_{0}, \mathcal{D}}\left[ \sum^{n_{\mathrm{vars}}}_{k=1}\sum^{n_{\mathrm{grid}}}_{l=1} \log\big(q_{\boldsymbol{\theta}}(z_{1, k, l}\mid\mathbf{z}_{\tau})\big)\right] + C\label{eq:vdm_loss}
\end{equation}
with $z_{1, k, l}$ as the terminal state for the $k$-th variable and $l$-th grid point.

Variational flow matching is a generalisation of the previously stated flow matching.
The linearity of the conditional velocity allows us to reduce the variational distribution to a mean-field approximation, where the distribution over all variables and grid points is factorised.
By combining mean-field variational flow matching with a Gaussian assumption, we recover conventional flow matching up to a constant \cite{eijkelboom_variational_2024},
\begin{equation}
    q_{\boldsymbol{\theta}}(\mathbf{z}_{1}\mid\mathbf{z}_{\tau}) = \mathcal{N}\left(\widehat{\boldsymbol{\mu}}_{\boldsymbol{\theta}}, (1-\tau)^2\mathbf{I}\right)\label{eq:vdm_gaussian_assumption}
\end{equation}
with
\begin{equation}
    \widehat{\boldsymbol{\mu}}_{\boldsymbol{\theta}}=\mathbf{z}_{\tau} + (1-\tau)\widehat{\mathbf{v}}_{\boldsymbol{\theta}}(\mathbf{z}, \tau, \mathbf{x}_{t}, \mathbf{F}_{t:t+12h}).\label{eq:vdm_mean_prediction}
\end{equation}
The covariance, $(1-\tau)^2\mathbf{I}$, is chosen such that the scaling of the mean-squared error vanishes,
\begin{align}
    \mathcal{L}_{\mathrm{VFM}}(\boldsymbol{\theta}) &= \mathbb{E}_{\tau, \mathbf{z}_{0}, \mathcal{D}}\left[\frac12\frac{1}{(1-\tau)^2} \lVert\boldsymbol{\mu}_{\mathrm{truth}}-\widehat{\boldsymbol{\mu}}_{\boldsymbol{\theta}}\lVert^2_{2}+C_{1}\right]\notag\\
    &= \mathbb{E}_{\tau, \mathbf{z}_{0}, \mathcal{D}}\left[\frac{1}{2(1-\tau)^2}\lVert (1-\tau)\mathbf{v}-(1-\tau)\widehat{\mathbf{v}}_{\boldsymbol{\theta}}(\mathbf{z}, \tau, \mathbf{x}_{t}, \mathbf{F}_{t:t+12h})\rVert^2_{2}\right] + C_{2}\notag\\
    &\propto \mathbb{E}_{\tau, \mathbf{z}_{0}, \mathcal{D}}\left[\lVert \mathbf{v}-\widehat{\mathbf{v}}_{\boldsymbol{\theta}}(\mathbf{z}, \tau, \mathbf{x}_{t}, \mathbf{F}_{t:t+12h})\rVert^2_{2}\right],
\end{align}
with $\boldsymbol{\mu}_{\mathrm{truth}}=\mathbf{z}_{\tau} + (1-\tau)\mathbf{v}$ and the constants $C_{1}$ and $C_{2}$, which do not depend on $\boldsymbol{\theta}$.
Hence, flow matching still adheres to a Gaussian assumption for the predictions, like the deterministic model.
Yet, its covariance diminishes through the successive refinements, yielding a sample from the targeted probability distribution.

An interesting case arises when we employ the model with a single step, mapping directly from $\tau=0$ to $\tau=1$: there, the latent state for $\tau=0$ contains no information about the prediction.
Consequently, if a neural network is trained for that, it would learn to disregard the latent state as input and to target a mean prediction \cite{liu_flow_2022}, similar to the deterministic model.
In this case, the model effectively reduces to a deterministic prediction model.

The weighting in the loss function for the deterministic model and flow matching is determined by the climatological standard deviation.
However, each of the six predicted variables is subject to distinct physical processes, leading to varying degrees of predictability, which might be not captured by their climatological weighting \cite{cipolla_multitask_2018}.

We account for this in a data-dependent weighting by altering the Gaussian assumption of variational flow matching,~Equation~\ref{eq:vdm_gaussian_assumption}, and making some covariances learnable.
As a simplification and exploiting the mean-field approximation, we introduce a per-variable standard deviation that depends on the pseudo time, $\mathbf{s}(\tau)$, obtaining
\begin{equation}
    q_{\boldsymbol{\theta}}(\mathbf{z}_{1}\mid\mathbf{z}_{\tau}) = \mathcal{N}\left(\widehat{\boldsymbol{\mu}}_{\boldsymbol{\theta}}, \frac12\Big((1-\tau)\cdot\mathbf{s}(\tau)\Big)^2\mathbf{I}\right),\label{eq:learnable_var_dist}
\end{equation}
as variational distribution.
The corresponding loss function for flow matching reads
\begin{equation}
    \mathcal{L}_{\textrm{VFM}}(\boldsymbol{\theta}, \mathbf{s}) = \mathbb{E}_{\tau, \mathbf{z}_{0},\mathcal{D}} \left[\frac{1}{2\cdot \big(\mathbf{s}(\tau)\big)^2}\lVert \mathbf{u}-\widehat{\mathbf{v}}_{\boldsymbol{\theta}}(\mathbf{z}, \tau, \mathbf{x}_{t}, \mathbf{F}_{t:t+12h})\rVert^2_{2}+\log\big(\mathbf{s}(\tau)\big)\right],
    \label{eq:learnable_loss_flow_matching}
\end{equation}
which has a similar formulation for diffusion models \cite{karras_analyzing_2023}.
The squared standard deviation, $\big(\mathbf{s}(\tau)\big)^2$, converges to half of the expected per-variable mean-squared error of the velocity and adapts automatically to the predictability of different variables.

Note that, while the weighting in the loss function is learned, the output of the neural network is still scaled by the climatological standard deviation.
This disentanglement between the scaling and the learned weighting becomes especially interesting when the neural network is trained across different datasets or resolutions: we can set a constant scaling of the neural network and learn different weightings per dataset (Fig. \ref{fig:app_weighting}a).
The weighting concentrates around the relevant scales, acting as gradient normalisation and stabilising the training \cite{karras_analyzing_2023}.

This learnable weighting can reduce the need to tune the scheduler during training and prediction.
We can also use the learned standard deviation to guide our inference scheduler (Fig. \ref{fig:app_weighting}b).
The larger the learned standard deviation, the larger the expected error for that scale, and importance sampling suggests that we should oversample these scales \cite{kingma_understanding_2023}.
Hence, we want to concentrate our pseudo time scheduler during prediction to the scales where the expected error is large, which are in our case the small scales.
While these small scales might be unimportant for perceptual generation like images, they can become especially important when the model is auto-regressively cycled, as errors at small scales can amplify over time and impact larger scales.
As prediction scheduler, we use a shifted sigmoid weighting, $\tau_{i} = \sigma(\alpha \cdot (\frac{i}{20}-0.5) + \beta)$ with $\alpha=3$ and $\beta=0.5$, which we normalise into the range $\tau_{0} = 0$ and $\tau_{1}=1$.
This scheduler shifts integration steps at intermediate scales towards smaller scales, increasing the importance therein.

Sea-ice states are inherently bounded: sea-ice and snow thickness are non-negative, while sea-ice concentration and damage are limited between 0 and 1.
Furthermore, the absence of sea ice (zero thickness) implies a lack of motion.
We hypothesise that explicitly incorporating these physical priors will improve the prediction accuracy.

Static thresholding, where the predictions are clipped into its physical bounds, can enforce these bounds \cite{saharia_photorealistic_2022}.
However, models trained with the mean-squared error are biased and prone to spurious fluctuations near these boundaries \cite{chen_analog_2023}.
These fluctuations can be particularly problematic in autoregressive predictions, where they may amplify instabilities in long-term simulations.
Generative flow models that incorporate bounds via reflection or log-barriers \cite{lou_reflected_2023,fishman_diffusion_2023} enforce a bound-avoiding behaviour such that the bound is never reached, leading to unphysical predictions, where, e.g., all grid points exhibit sea ice, similar to a truncated distribution approximation.
Instead, we introduce censored flow matching, accounting for static thresholding during training.

\begin{figure}[ht]
    \centering
    \includegraphics{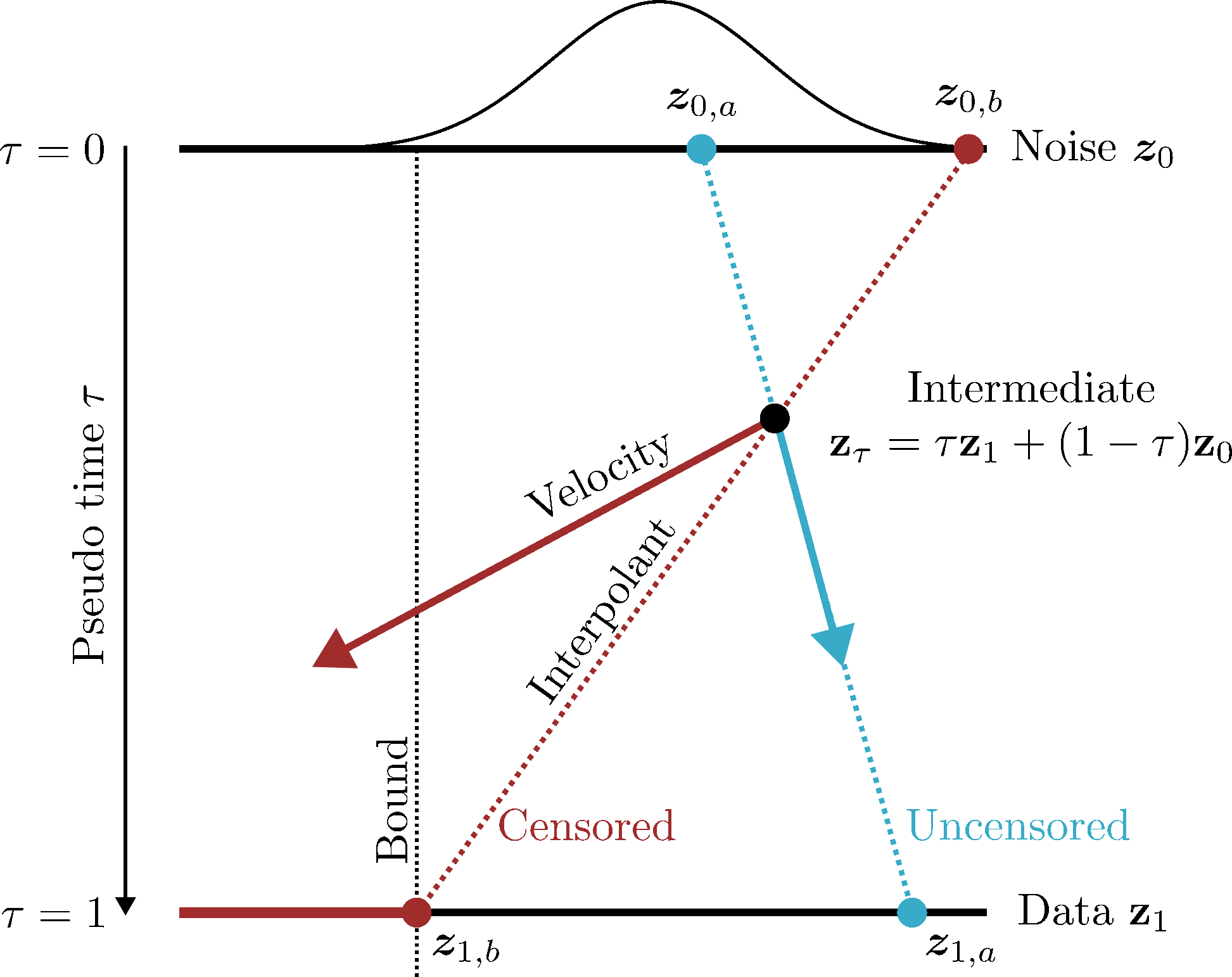}
    \caption{
        \textbf{Schematic showing the principle of censored flow matching.}
        For two data cases (a and b), we draw independent noise during training and estimate a linear interpolant (dotted lines), connect the data and the noise.
        For a (blue), the data sample, $z_{1, a}$, is not on the bound (uncensored) and the targetted velocity points toward the data.
        For b (red), the data sample, $z_{1, b}$, is on the bound and, hence, censored; the true data value is unknown, it is only known that it lies between $-\infty$ and the bound as indicated by the red horizontal line.
        Here, the targetted velocity points towards an increase in the probability that the recovered data exceeds the bound.
    }\label{fig:app_censored_schematic}
\end{figure}

In censoring, all predictions that exceed a bound are exactly mapped to that bound.
We generally assume that there exist true sea-ice dynamics without physical bounds, e.g., in the tropics, sea ice would still melt even though no sea ice is observed there.
When the observed data is censored to a bound, its dynamics are unknown, and we have just the information that the dynamics exceeded the bound.
In this case, the approximated velocity should increase the probability that the reproduced prediction exceeds the bound as well.
To achieve this, we use the predicted velocity to estimate an approximated cumulative distribution function (CDF) from the variational Gaussian distribution.
The CDF is then evaluated at the bounded velocity, and its logarithm replaces the mean-squared error loss for censored data points.
A visualisation of the targetted training process is shown in Fig. \ref{fig:app_censored_schematic}.

In this framework, the variational Gaussian distribution,~Equation~\ref{eq:learnable_var_dist}, defines the latent distribution \textbf{before} the terminal state has been censored,
\begin{equation}
    q_{\boldsymbol{\theta}}(\widetilde{\mathbf{z}}_{1} \mid \mathbf{z}_{\tau}) = \mathcal{N}\left(\widehat{\boldsymbol{\mu}}_{\boldsymbol{\theta}}, \frac12\Big((1-\tau)\cdot\mathbf{s}(\tau)\Big)^2\mathbf{I}\right).
\end{equation}
To censor the terminal state, we have to define a thresholding function that depends on the physical lower $x^{L}_{k}$ and upper bound $x^{U}_{k}$, here for the $k$-th variable.
The terminal state for each of the $k$ variables and each of the $l$ grid points is independently censored by using its latent bounds $z^{L}_{k, l} = (x^{L}_{k}-x_{t,  k, l})/{\sigma_{\mathrm{clim}, k}}$ and $z^{U}_{k, l} = (x^{U}_{k}-x_{t,  k, l})/{\sigma_{\mathrm{clim}, k}}$, respectively.
The thresholding function reads
\begin{equation}
    f(\widetilde{z}_{1, k, l}) = \begin{cases}
        z^{L}_{k, l}, &\text{if } \widetilde{z}_{1, k, l} < z^{L}_{k, l} \\
        z^{U}_{k, l}, &\text{if } \widetilde{z}_{1, k, l} > z^{U}_{k, l} \\
        \widetilde{z}_{1, k, l}, & \text{otherwise}.
    \end{cases}
\end{equation}

Given a Dirac delta distribution centred around this thresholding function, ${p(z_{1, k, l} \mid \widetilde{z}_{1, k, l}) = \delta\left(z_{1, k, l} - f(\widetilde{z}_{1, k, l})\right)}$, we define the probability density function (PDF) of the censored Gaussian distribution,
\begin{equation}
    q_{\boldsymbol{\theta}}(z_{1, k, l} \mid \mathbf{z}_{\tau}) =\begin{cases}
    \frac{1}{s_{k}}\varphi(\frac{z_{1, k, l}-\mu_{\boldsymbol{\theta}, k, l}}{s_{k}}), & z^{L}_{k, l} < \widetilde{z}_{1, k, l} < z^{U}_{k, l}\\
    \Phi(\frac{z^{L}_{k, l}-\mu_{\boldsymbol{\theta}, k, l}}{s_{k}}), &\text{if } \widetilde{z}_{1, k, l} = z^{L}_{k, l}\\
    \Phi(\frac{\mu_{\boldsymbol{\theta}, k, l}-z^{U}_{k, l}}{s_{k}}), &\text{if } \widetilde{z}_{1, k, l} = z^{U}_{k, l}\\
    0, & \text{otherwise},
    \end{cases}
\end{equation}
with $\varphi(\cdot)$ as the PDF and $\Phi(\cdot)$ as the cumulative density function of a Gaussian distribution with mean $0$ and standard deviation $1$.
This PDF of a censored Gaussian distribution basically maps all the density that exceeds the bound to exactly the bound, while the density between the bounds remains as usual.

Using this variational distribution in the variational flow matching framework, we obtain as loss function for censored flow matching,
\begin{align}
    J(\mu_{\boldsymbol{\theta}, k, l}, &s_{k}, z_{1, k, l}, z^{L}_{k, l}, z^{U}_{k, l})\notag\\
    =~&\mathrm{I}(z^{L}_{k, l} < \widetilde{z}_{1, k, l} < z^{U}_{k, l})\left[ \frac{1}{2\cdot(s_k)^2}(u_{k, l}-\widehat{v}_{\boldsymbol{\theta}, k, l})^2 + \log(s_{k})\right]\notag\\
    &-\mathrm{I}(\widetilde{z}_{1, k, l} = z^{L}_{k, l})\left[ \log\Phi\Big(\frac{u_{k, l}-\widehat{v}_{\boldsymbol{\theta}, k, l}}{s_k}\Big)\right]\notag\\
    &-\mathrm{I}(\widetilde{z}_{1, k, l} = z^{U}_{k, l})\left[ \log\Phi\Big(\frac{\widehat{v}_{\boldsymbol{\theta}, k, l}-u_{k, l}}{s_k}\Big)\right],\label{eq:censor_fm_loss}
\end{align}
with    
\begin{equation}
    \mathcal{L}_{\mathrm{GenSIM}}(\boldsymbol{\theta}, \mathbf{s}) = \mathbb{E}_{\tau, \mathbf{z}_{0}, \mathcal{D}}\left[ \sum^{n_{\mathrm{vars}}}_{k=1}\sum^{n_{\mathrm{grid}}}_{l=1} J(\mu_{\boldsymbol{\theta}, k, l}, s_{k}, z_{1, k, l}, z^{L}_{k, l}, z^{U}_{k, l})\right],
\end{equation}
and $\mathrm{I}(\cdot)$ is the indicator function.
The predicted velocity is no longer only used to estimate the dynamics but also to estimate the probability that the bound is exceeded \cite{finn_diffusion_2024};
the neural network is trained in a mixed classification-regression setting, also called Tobit model \cite{tobin_estimation_1958}.
In settings with physical bounds, the resulting estimator for the velocity is consistent and gives an unbiased estimate of the true but only partially observed velocity \cite{amemiya_regression_1973}.
As schematically shown in Fig. \ref{fig:app_censored_schematic}, the loss function pushes the velocity towards a prediction that exceeds the bound, when the observed data is exactly on the bound, while the velocity in the uncensored case remains the same.

To use the censored estimate in the sampler, we have to reduce the distribution to a point estimate, accounting for the censoring.
Here, we treat the neural network prediction as point estimate that specifies the median of the velocity.
At each integration step, we project with the predicted velocity the intermediate state to the terminal state and apply thresholding when the projected state exceeds the bound,
\begin{equation}
    \widehat{\mathbf{z}}_{1 \mid \tau} = \mathrm{minimum}\Big(\mathrm{maximum}\big(\mathbf{z}_{\tau} + \tau \widehat{\mathbf{v}}_{\boldsymbol{\theta}}, (\mathbf{\mathbf{x}^{L}}-\mathbf{x}_{t})/\boldsymbol{\sigma}_{\mathrm{clim}}\big), (\mathbf{\mathbf{x}^{U}}-\mathbf{x}_{t})/\boldsymbol{\sigma}_{\mathrm{clim}}\Big)
\end{equation}
with $\mathrm{minimum}(\cdot, \cdot)$ and $\mathrm{maximum}(\cdot, \cdot)$ as minimum and maximum function, respectively.
By applying~Equation~\ref{eq:vfm_cond_velocity}, we obtain the velocity from the potentially thresholded projected state,
\begin{equation}
    \widehat{\mathbf{v}}_{\boldsymbol{\theta}} = \frac{ \widehat{\mathbf{z}}_{1 \mid \tau} - \mathbf{z}_{\tau}}{1-\tau}.
\end{equation}
This median estimate of the \textit{thresholded} velocity provides a simple yet efficient way to produce samples that can remain directly on the bounds without introducing additional randomness as needed in a Monte-Carlo sampling scheme.

In preliminary experiments, we found that censored flow matching boosts the prediction performance compared to conventional flow matching, confirming that incorporating prior physical knowledge can improve generative modelling.
The variational flow matching thereby paves a general way to incorporate prior physical knowledge into flow matching.

\subsubsection*{Transformer architecture}\label{sec:app_arch_trans}

The velocity for the generative flow is approximated by a neural network, which is trained to work on subdomains of size $80 \times 80$.
Our curvilinear mesh is a structured mesh with $512 \times 512$ non-Cartesian cells, which are partially masked due to land.
Accounting for these considerations, we built our neural network around a diffusion transformer architecture \cite{peebles_scalable_2023}, inspired by design choices from large language models and from recent efficient diffusion models \cite{crowson_scalable_2024}.

The neural network takes as main input $24$ channels--consisting of the six state variables $\mathbf{z}_{\tau}$, the six sea-ice state variables $\mathbf{x}_{t}$, and the twelve atmospheric forcing variables $\mathbf{F}_{t:t+12\,\textrm{h}}$ (four variables for two time steps and four channels for the degree day features).
The transformer needs as additional input the mesh in Cartesian coordinates in a north polar stereographic projection, and the land mask ($1=\text{ocean}$, $0=\text{land}$).
The scalar input that is converted into an embedding is comprised of the pseudo time $\tau \in [0, 1]$, the resolution in kilometer, and three labels for the distribution augmentation \cite{jun_distribution_2020, finn_generative_2024}.

Instead of mesh-based cell values, the transformer processes flattened word-like tokens without any ordering.
Furthermore, due to self attention, the transformer scales quadratically with the number of tokens.
Hence, we have to project the raw input cells into aggregated tokens through tokenisation.

Prior to tokenisation, all values on land are set to $0$ and the land mask is added as additional channel to achieve a constant bias-like feature across unmasked cells.
The tokenisation is a learnable linear projection without bias, which combines $2\times2$ cells into a token, reducing the data from $80 \times 80$ cells to $1600$ tokens.
To counter-act the data compression, the layers inflate the number of channels from $24$ to $512$.
The mesh is pooled by the average in the $2\times2$ window, while the mask is maximum pooled.

The scalar inputs are converted into an embedding concurrently to the input tokenisation.
The pseudo time, resolution, and labels are independently processed by combining random Fourier features (RFF)\cite{rahimi_random_2007}, which expand the number of features to $512$, with a linear layer, a Sigmoid Linear Unit (SiLU) activation function \cite{hendrycks_gaussian_2016}, and another linear layer, keeping the number of feature at $512$.
At the end, each of the three branches is added together and activated with a SiLU to obtain the embedding.

The tokens are processed in a sequence of $8$ transformer blocks, in which self-attention blocks are interleaved with small multi-layer perceptrons (MLP).
Each of the self-attention blocks and MLPs is added to a short skip connection that directly links the input to the output of each block \cite{he_deep_2015}.
All self-attention blocks have the same structure:
\begin{itemize}
    \item {
        The input is normalised by root-mean-squared normalisation (RMSNorm) \cite{zhang_root_2019} where the dimension-wise scaling coefficient is a linear combination of the embedding \cite{peebles_scalable_2023}.
    }
    \item{
        The normalised tokens are converted into a queries, keys, and values by a linear projection without bias, each containing $8$ heads and $64$ features per head and per token.
    }
    \item{
        The queries and keys are normalised by RMSNorm with a learnable scaling parameter along the feature dimension.
    }
    \item{
        The queries and keys are rotated by a two-dimensional rotary positional embedding (RoPE) \cite{su_roformer_2024, heo_rotary_2024} based on the tokenised mesh and scaled by the resolution.
        This efficiently implement an anisotropic relative distance dependency into the self-attention block.
    }
    \item{
        After applying RoPE, multi-head self attention \cite{vaswani_attention_2017} is applied along the feature dimension, extracting $8$ different spatial features per block.
    }
    \item{
        A linear projection without bias transforms the self-attention out back into $512$ features per token.
    }
\end{itemize}
Before added to the short-skipped tokens, the output of each self-attention block is multiplied by the tokenised mask to ensure that the land-masked tokens are always zero.

After adding the self-attention output to the tokens, an MLP is applied which has the following structure:
\begin{itemize}
    \item Like for the self-attention block, the input to the MLP is normalised by RMSNorm and scaled dependent on the embedding.
    \item Afterwards, a linear layer without bias doubles the number of features per token to $1024$.
    \item A gated SiLU activation function \cite{shazeer_glu_2020} activates the inflated features, before the output is deflated once again to $512$ features by a linear projection without bias.
\end{itemize}
The output, multiplied by the mask, is added to the short-skipped tokens, like for the self-attention block.

To stabilise the transformer by imitating a U-Net-like architecture \cite{ronneberger_unet_2015, chen_stabilized_2025}, we implement long-skip connections, linking the output of the tokeniser to the input for the final head, linking the output of the first transformer block to the input of the last transformer block, and so forth.
As linking function, we use a linear combination \cite{crowson_scalable_2024}, which is dimension-wise modulated by the embedding.

In the end, also called the head, the tokens are normalised by RMSNorm and scaled with the embedding.
To extract features that can contain sharp edges, we apply a rectified linear unit (ReLU) activation function after the normalisation.
In the end, we linearly project the $512$ features per token to $24$ output features ($6 \cdot 2\times2$).
Applying pixel shuffle \cite{shi_realtime_2016}, the output features are rearranged to form $6$ predicted velocity values per cell.

The standard deviation $\mathbf{s}$, used in the censored flow matching model, is produced by a similar neural network to the embedding, combining RFFs with linear layers and a SiLU activation function.
After adding the branches together and activating with a SiLU activation function, the output is linearly combined to $6$ output features with a learnable bias term.
To avoid negative standard deviations, we parametrise the output as its logarithm, $\log(\mathbf{s})$.

The transformer has in total $28.1\times10^6$ parameters, which is smaller than typical generative models for image generation \cite{crowson_scalable_2024} and for weather prediction \cite{couairon_archesweather_2024}.
In early experiments, we have seen that we could increase the number of layers and the feature dimensionality to around $500\times10^6$ parameters without overfitting to improve the scores.
Yet, for efficiency, we constrain the number of parameters.
The model for the standard deviation adds $1.6\times10^6$ parameters to the training of the generative flow.






\begin{figure}[ht]
    \centering
    \includegraphics[width=0.8\textwidth]{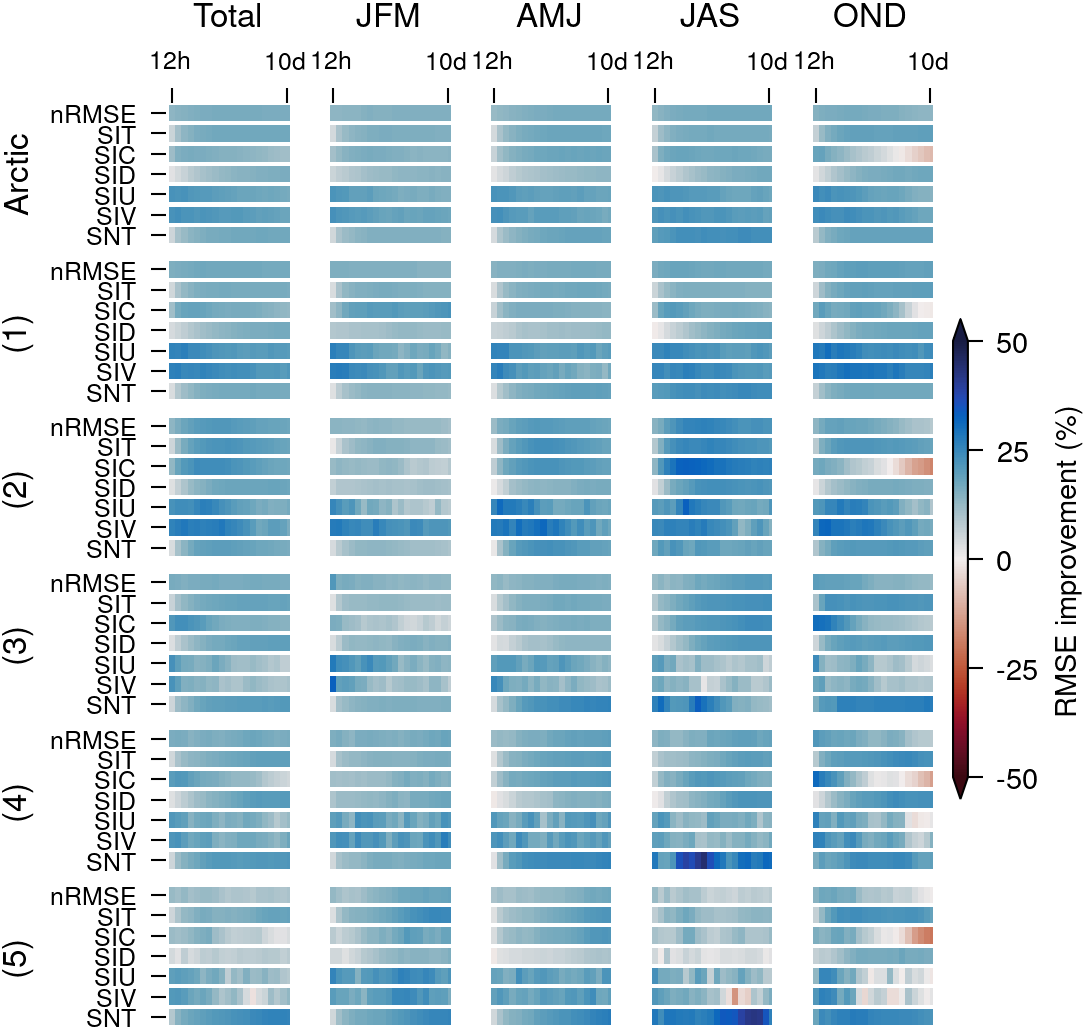}
    \caption{
        \textbf{GenSIM's ensemble mean has a lower RMSE than a deterministic baseline across regions, variables, and seasons.}The score card, depicting the relative root-mean-squared error (RMSE) of GenSIM's ensemble mean to the RMSE of the deterministic baseline for different regions, variables, and seasons, as estimated over all ice-covered grid points (sit=thickness, sic=concentration, sid=damage, siu=$u$-drift, siv=$v$-drift, snt=snow-on-ice thickness).
        A blue RMSE corresponds to an improvement of GenSIM, while a red one shows a degradation of the forecast.
        The regions (1--5) are depicted and explained in Fig. \ref{fig:app_regions} and remaining regions are shown in Fig. \ref{fig:app_scorecard_b}.
    }\label{fig:app_scorecard_a}
\end{figure}

\begin{figure}[ht]
    \centering
    \includegraphics[width=0.8\textwidth]{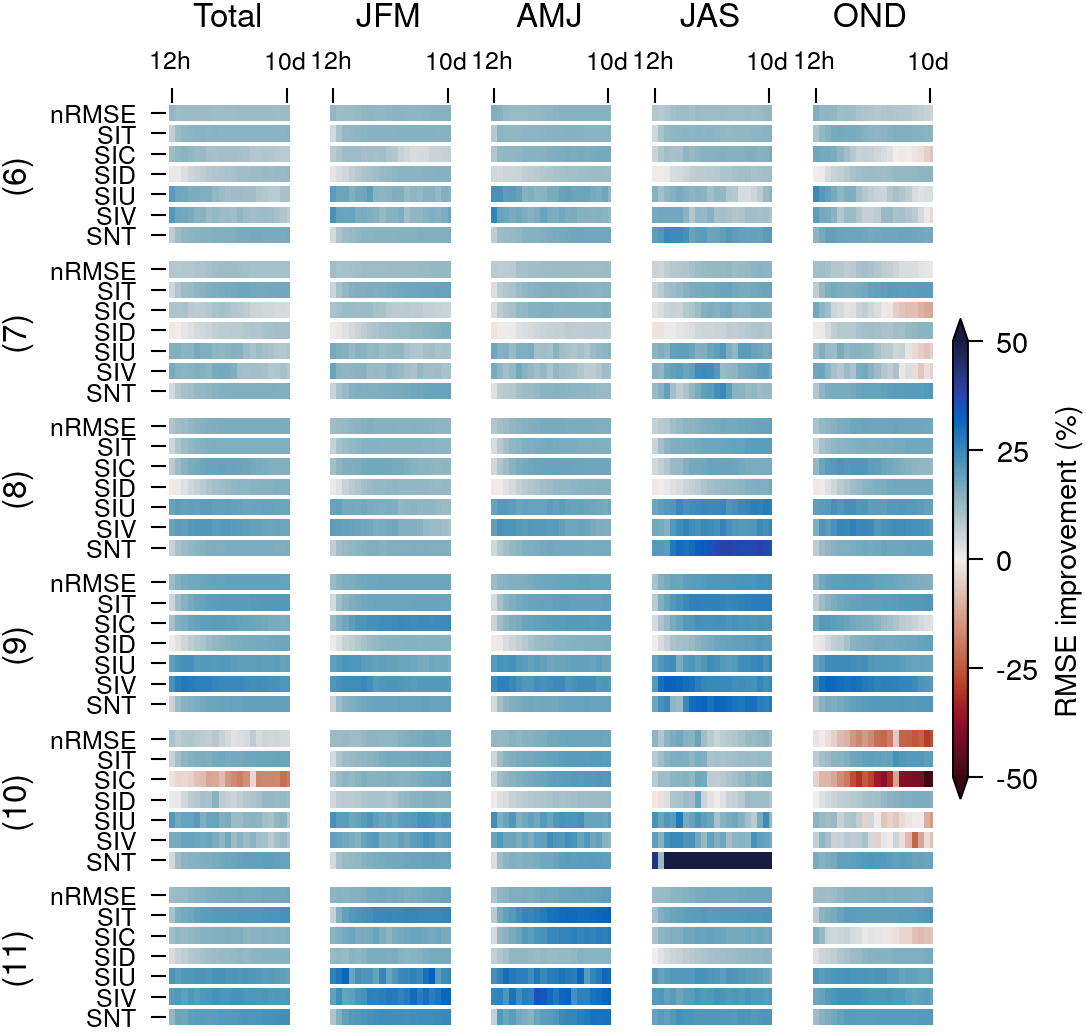}
    \caption{
        Continuation of Fig. \ref{fig:app_scorecard_a} for regions (6--11).
    }\label{fig:app_scorecard_b}
\end{figure}

\clearpage 





\end{document}